\documentclass[
  journal=pasa,
  manuscript=research-paper, %% or "review"
  year=2024,
  volume=x,
]{cup-journal-1}

\usepackage{graphicx}
\usepackage{longtable}
\usepackage{lmodern}

\usepackage{scalerel}
\usepackage{tikz}
\usetikzlibrary{svg.path}

\definecolor{orcidlogocol}{HTML}{A6CE39}
\tikzset{
  orcidlogo/.pic={
    \fill[orcidlogocol] svg{M256,128c0,70.7-57.3,128-128,128C57.3,256,0,198.7,0,128C0,57.3,57.3,0,128,0C198.7,0,256,57.3,256,128z};
    \fill[white] svg{M86.3,186.2H70.9V79.1h15.4v48.4V186.2z}
                 svg{M108.9,79.1h41.6c39.6,0,57,28.3,57,53.6c0,27.5-21.5,53.6-56.8,53.6h-41.8V79.1z M124.3,172.4h24.5c34.9,0,42.9-26.5,42.9-39.7c0-21.5-13.7-39.7-43.7-39.7h-23.7V172.4z}
                 svg{M88.7,56.8c0,5.5-4.5,10.1-10.1,10.1c-5.6,0-10.1-4.6-10.1-10.1c0-5.6,4.5-10.1,10.1-10.1C84.2,46.7,88.7,51.3,88.7,56.8z};
  }
}
\newcommand\orcidicon[1]{\href{https://orcid.org/#1}{\mbox{\scalerel*{
\begin{tikzpicture}[yscale=-1,transform shape]
\pic{orcidlogo};
\end{tikzpicture}
}{|}}}}

\title{Investigating Unusual H$\alpha$ Features towards the Scutum Supershell}

\author{R.~Alsulami}
\affiliation{School of Physics, Chemistry and Earth Sciences, The University of Adelaide, Adelaide SA 5005, Australia}
\alsoaffiliation{Astronomy Dept, Faculty of Science, King Abdulaziz University, Jeddah, Saudi Arabia}

\email[R.~Alsulami \orcidicon{0000-0002-4934-7422} ]{rami.alsulami@adelaide.edu.au }

\author{S.~Einecke  \orcidicon{0000-0001-9687-8237}}
\affiliation{School of Physics, Chemistry and Earth Sciences, The University of Adelaide, Adelaide SA 5005, Australia}

\author{G.~P.~Rowell \orcidicon{0000-0002-9516-1581}}
\affiliation{School of Physics, Chemistry and Earth Sciences, The University of Adelaide, Adelaide SA 5005, Australia}

\author{P.~K.~McGee \orcidicon{0000-0002-9576-442X}}
\affiliation{School of Physics, Chemistry and Earth Sciences, The University of Adelaide, Adelaide SA 5005, Australia}

\author{M.~D.~Filipovi\'c \orcidicon{0000-0002-4990-9288}}
\affiliation{Western Sydney University, Locked Bag 1797, Penrith South DC, NSW 2751, Australia}

\author{I.~R.~Seitenzahl \orcidicon{0000-0002-5044-2988}}
\affiliation{Heidelberg Institute for Theoretical Studies, Schloss-Wolfsbrunnenweg 35, 69118 Heidelberg, Germany}

\author{M.~Stupar\, \orcidicon{0000-0002-0338-9539}}
\affiliation{Western Sydney University, Locked Bag 1797, Penrith South DC, NSW 2751, Australia}

\author{T.~Collins \orcidicon{0000-0001-5020-5387} }
\affiliation{School of Physics, Chemistry and Earth Sciences, The University of Adelaide, Adelaide SA 5005, Australia}
\alsoaffiliation{Universit\"at Potsdam, Institut f\"ur Physik und Astronomie, Campus Golm, Haus 28, Karl-Liebknecht-Str. 24/25, 14476 Potsdam-Golm, Germany}

\author{Y. Fukui \orcidicon{0000-0002-8966-9856}}
\affiliation{Department of Physics,  Nagoya University, Furo-cho, Chikusa-ku, Nagoya 464-8601, Japan}

\author{H. Sano \orcidicon{0000-0003-2062-5692}}
\affiliation{Faculty of Engineering, Gifu University, Yanagido 1-1, Gifu, 501-1193, Japan}

\received {03 08 2023}
\revised  {18 05 2024}
\accepted {17 10 2024}
\published{dd Mmm YYYY}

\usepackage{hyperref}
\hypersetup{colorlinks,citecolor=blue,linkcolor=blue,urlcolor=blue}
\usepackage[normalem]{ulem}
\usepackage{booktabs}
\usepackage{amssymb}

\newcommand{\PP}{HESS\,J1825$-$137}
\newcommand{\PPP}{HESS\,J1826$-$130}

\newcommand{\blue}{\textcolor{black}}

\newcommand{\HI}{\mbox{H\scriptsize{\sc\,I}}}
\newcommand{\HII}{\mbox{H\scriptsize{\sc\,II}}}
\usepackage{pdflscape}
\usepackage{soul}

\newcommand{\serob}{Ser\,OB1B}

\keywords{H$\alpha$ -- OB association -- Early-type stars -- HI shells -- Photo-ionisation -- ISM: jets and outflows}

\begin{document}

\begin{abstract}
    We investigate the unusual H$\alpha$ features found towards the Scutum Supershell via recent arc-minute and arc-second resolution imaging. These multi-degree features resemble a long central spine ending in a bow-shock morphology. We performed a multi--wavelength study in [S\,II] optical, radio continuum, infrared continuum, \HI{}, CO, X-ray and gamma-ray emissions. Interestingly, we found the Galactic worm GW\,16.9$-$3.8 \HI{} feature appears within the Scutum Supershell, and  likely influences the spine morphology. {Furthermore,
     the rightmost edge of the bow-shock H$\alpha$ emission  overlaps with [S II] line emission,  4.85 GHz radio, and both 60$\mu$m and 100$\mu$m infrared continuum emissions, suggesting some potential for excitation by shock heating.} 
    We estimated the photo-ionisation from O-type and B-type stars in the region (including those from the OB associations Ser\,OB1B, Ser\,OB2 and Sct\,OB3) and found that this mechanism could supply the excitation to account for the observed H$\alpha$ luminosity of the spine and bow-shock of $\sim 1 - 2 \times 10^{36}$\,erg\,s$^{-1} (d/2.5\,{\rm kpc})^2$. 
    \blue{Recent MHD simulations by \citet{2022OAst...31..154D} demonstrate the potential for supernova events to drive outflow and bow-shock types of features of the same energetic nature and physical scale as the H$\alpha$ emission we observe here. 
    While this clearly requires many supernova events over time, we speculate that one contributing event could have come from the presumably energetic supernova (hypernova) birth of the magnetar tentatively identified in the X-ray binary LS 5039.
     }
\end{abstract}

\section{Introduction} \label{sec:intro}
Supershells { - expanding cavities in neutral hydrogen -} are highly energetic features in the interstellar medium (ISM), which extend to hundreds of parsecs away from the Galactic plane.
The formation of a supershell can arise from stellar winds of O- or B-type stars \citep{1979ApJ...229..533H}, with additional energy coming from supernova or hypernova events from the core collapse of these massive stars \citep{2000ApJ...539..706P}.
These processes can push the neutral ISM away from the Galactic plane (in particular the atomic gas traced by the 21\,cm \HI{} line). 
{As a result,} long `chimney' \HI{} features, possibly enhanced by Rayleigh-Taylor instabilities, can extend beyond the Galactic plane. 
Related linear `worm' features {in the \HI{} gas} may also form as the supershell breaks up  %\red{due to specific events, for example, supernovae} 
\citep[see e.g.][]{2001MNRAS.328..708D,2013PASA...30...25D}. 
Ionised gas, traced by the optical H$\alpha$ line, has also been associated with supershells \citep{1997ApJ...474L..31D,2012A&A...543A..26M}.
The Milky Way is host to several dozen known \HI{} supershells with mechanical energies of $>10^{52}$\,erg (see \autoref{tab:obGSH} for a list of Galactic supershells including their estimated energy and distance). 

The Scutum Supershell GSH\,018$-$04+44  \citep{1979ApJ...229..533H,2000A&AS..143...33B} is one of the most energetic supershells. 
\citet{2000ApJ...532..943C} provided the most detailed study of this supershell so far, estimating its distance at 3.3\,kpc with a physical extension of approximately 290\,pc based on its \HI{} morphology. 
The supershell has an \HI{} density of about 4\,/\,cm$^{3}$ and a total energy of 1.1$\times$10$^{52}$\,erg. 
Several supernova remnants (SNRs) and \HII{} regions were identified around or within the supershell.

Multiple OB associations are present in this region, which can host supernova events that feed energy into the supershell. 
The catalogue of OB associations by \citet{1989AJ.....98.1598B} lists Ser\,OB and Sct\,OB3 within a few degrees north of the Scutum Supershell.
A later catalogue by \citet{1995Ast} suggests that Ser\,OB is actually three separate OB associations (Ser\,OB1A, \serob{}, and Ser\,OB2).
\citet{1992ApJ...390..108K} have suggested that these associations are potential energy sources of the Scutum Supershell. 

Based on observations from the Wisconsin H$\alpha$ Mapper~(WHAM) survey, \citet{2000ApJ...532..943C} also revealed an `optical blowout' phenomenon. 
They suggested that this was generated by a combination of SNRs, strong winds from O and B stars, and \HII{} regions.
{They also noted the presence of {hot} ionised gas within the supershell, based on the presence of thermal X-ray emission that spatially anti-correlates with the associated \HI{} gas.} 

{Several Galactic supershells feature optical H$\alpha$ emission and associated filaments.}
For instance, GSH\,305+01-24 is an energetic supershell with H$\alpha$ emission detected towards the inner side of its \HI{} shell \citep{2014A&A...562A..69K}. 
Other examples, featuring both H$\alpha$ and \HI{} emission, are the supershells GSH\,006-15+7 \citep{2012MNRAS.421.3159M}, {the Carina Flare} \citep{1999PASJ...51..751F,2008MNRAS.387...31D}, and  the W4 { \HII{} region and its} Superbubble \citep{1997ApJ...474L..31D,2009ApJ...691.1109L}.
In general, it is assumed that the H$\alpha$ emission, tracing the warm ionised medium, is formed due to ionisation by the stellar photons and/or  winds from O and B stars (e.g.\ \citet{2006ApJ...652..401M, 2013PASA...30...25D}). 

Since the study by \citet{2000ApJ...532..943C}, improved-quality (in terms of resolution and sensitivity) {and new data in the radio (\HI{} and continuum), infrared, optical (H$\alpha$ and S[II] spectral lines), and X-ray bands have become available.} 
Additionally, the population of energetic sources towards the Scutum Supershell (besides the O and B stars and SNRs discussed by \citet{2000ApJ...532..943C}) now includes a number of gamma-ray sources detected in the GeV to TeV energy bands, such as the pulsar wind nebulae \PP\ and \PPP\  \cite[and references within]{2018A&A...612A...1H,2021MNRAS.504.1840C}, plus the High-Mass X-ray Binary LS\,5039 \citep{2006A&A...460..743A} {which has been suggested to be a magnetar \citep{2020PhRvL.125k1103Y}.}

Intriguingly, a part of the extended H$\alpha$ emission in this region was attributed to a shock from SNR G18.7--2.2 based on the [S\,II]/H$\alpha$ emission ratio \citep{stupar2008newly}. 
This SNR was then postulated by \citet{voisin2016ism} to have formed from the progenitor supernova event that also gave birth to the pulsar powering the prominent TeV-bright pulsar wind nebula (PWN) \PP{}.

Given the higher-quality and additional observations available, plus the discovery of several high-energy sources towards the Scutum Supershell \citep{2018A&A...612A...1H}, a new investigation of the H$\alpha$ features towards the Scutum Supershell is warranted. 
\section{Data} 
\label{sec:data}

We made use of publicly available data across the wavebands of interest: H$\alpha$, \HI{}, infrared continuum, $^{12}$CO, radio continuum, X-rays and gamma rays. 
{Additionally, we have performed new H$\alpha$ and S[II] emission line observations in an effort to further characterise the origin of the H$\alpha$ emission.}

The full-sky H$\alpha$ map by \citet{finkbeiner2003full} is a composite of three surveys: SHASSA \citep{2001PASP..113.1326G}, VTSS \citep{1998PASA...15..147D} and WHAM \citep{haffner2003wisconsin}. 
This composite map features an angular resolution of 6\,arcmin. 
The WHAM observations by themselves provide H$\alpha$ intensities for different velocities (with respect to the local standard of rest) with a resolution of $12$\,km\,/\,s, and covering a range from -150 to 150\,km\,s$^{-1}$ \citep{1998PASA...15...14R}.
We also made use of the SuperCOSMOS H$\alpha$ Survey (SHS) with the UK Schmidt Telescope (UKST) of the Anglo-Australian Observatory (AAO/UKST) with an angular resolution of 1-2\,arcsec \citep{2005MNRAS.362..689P}.

Our dedicated observations of H$\alpha$ and [S\,II] utilised a Skywatcher Evolux 62ED refractor operating at f/5.6 with a reducer/corrector lens, a ZWO {ASI2600MM-Pro} camera with $3\times 3$ pixel binning (resulting in a resolution of $\sim$6\,arcsec per effective pixel), an Antlia 30\AA{} H$\alpha$ filter (transmission 90\% at 656.3\,nm; 38\% at [N\,II] 654.8\,nm, 25\% at 658.3\,nm), an Optolong 65\AA{} [S\,II] filter (transmission  94\% at [S\,II] 671.6\,nm, and 96\% at [S\,II] 673.1\,nm) and a broadband R filter.
Therefore, there will be some contamination from the [N\,II] lines in the H$\alpha$ observations, and both [S\,II] lines are recorded for the [S\,II] observations.
The [S\,II] images were obtained over two nights, one in September 2022, and one in May 2023. 
For the 2022 observation, the exposure was 120\,s, with 32 individual [S\,II] images being recorded. 
Short-exposure broadband R images were also recorded for continuum subtraction.
During the 2023 observation, twelve 300\,s [S\,II] exposures were taken, as well as a short-exposure R-band image. 
Thus, a total exposure time of the the [S\,II] observations is 7440\,s. 
Dark frames (to account for CCD noise) and flat-field frames (to correct uneven illumination) were obtained for each observing session. 
Intensity-scaled R-band images were subtracted from each night's stacked [S\,II] images. 
Image smoothing and bright/dark outlier reduction were applied in \texttt{AstroImageJ} \citep{2017AJ....153...77C} to reduce the noise in the image produced by imperfect continuum subtraction . 
Some inhomogeneities in the [S\,II] image remain, following flat-fielding and continuum subtraction. 
These are visible as large-angular-scale variations in brightness in the background. 
There are clear structures in the [S\,II] emission which are seen also in the H$\alpha$ observations, so these imperfections do not negate the detection of the stronger [S\,II] features found in these observations.
H$\alpha$ imagery was obtained in July 2023, with 235\,min of exposure time in a single night. 
A short R-band image was used for continuum subtraction. 
All intensities presented for our own H$\alpha$ and [S\,II] imagery are arbitrary, as no standard sources were observed, and no flux calibration was applied.

For the \HI{} gas, we made use of the Southern Galactic Plane Survey (SGPS) from the Parkes telescope \citep{2005ApJS..158..178M}, which extends to latitudes $|b|\leq 10^\circ$ and covers most of the Scutum Supershell region. 
The spatial resolution (16\,arcmin) and sensitivity (0.2\,K per channel) of the SGPS data are better than those of the \HI{} data used by \citet{2000ApJ...532..943C} (21\,arcmin and $\sim0.6$\,K per channel), but have similar spectral resolution (0.8\,km\,/\,s). 

\begin{figure*}
  \centering
   \includegraphics[width=\textwidth]{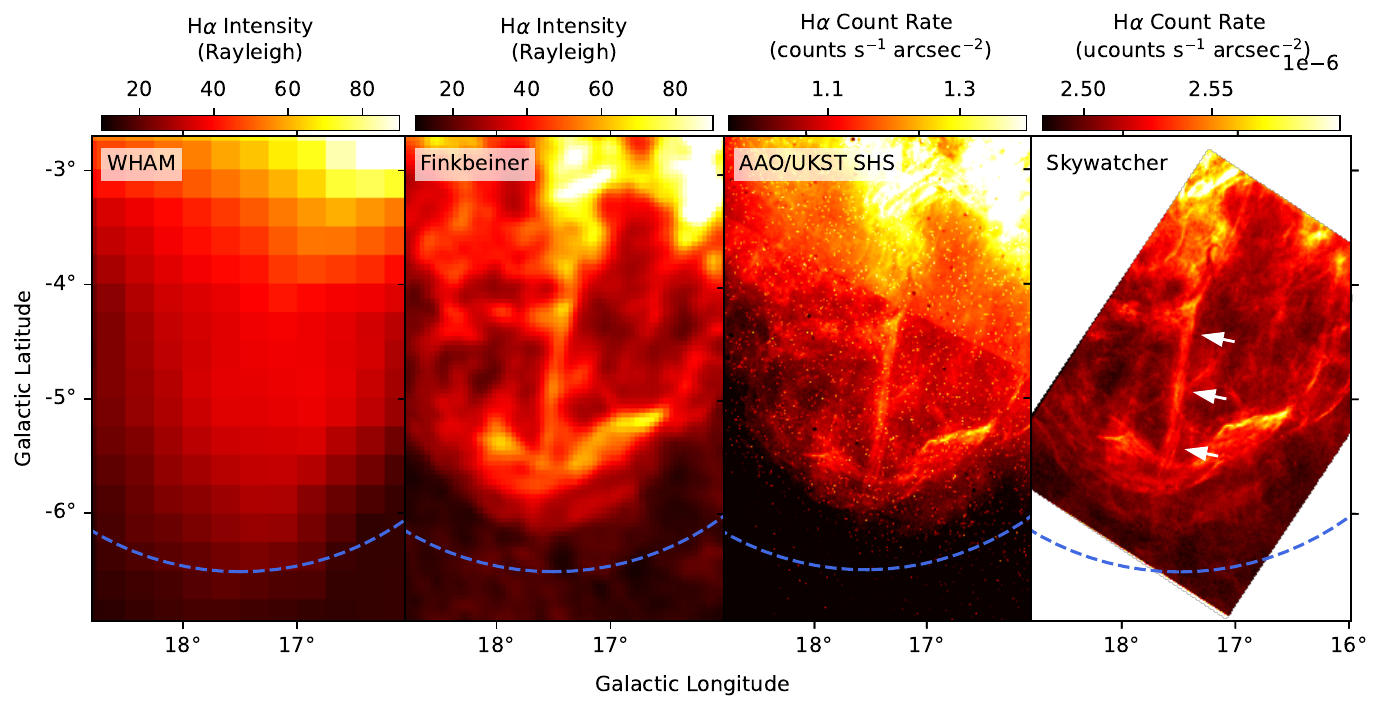}
  \caption{Comparison of different observations of H$\alpha$ emission towards the Scutum Supershell (dashed blue circle). The observations refer to the WHAM survey \citep{haffner2003wisconsin}, the composite map derived by \citet{finkbeiner2003full}, the AAO/UKST SHS survey \citep{2005MNRAS.362..689P}, and our own ({uncalibrated counts labeled as "ucounts"}) observations with the Skywatcher Evolux 62ED refractor with an H$\alpha$ filter.
    The `{spine}' feature has a length of at least 2$^\circ$ and is highlighted by white arrows.
    The blowout morphology surrounding the {spine} is clearly revealed in the higher-resolution images. The {bow-shock} feature at the southern end of the {spine is also evident in these images}.
    }  \label{fig:evo_h}
\end{figure*}

For the molecular hydrogen (H$_2$) gas, we made use of the CfA-Chile combined carbon monoxide $^{12}$CO(J=1--0) Galactic plane survey \citep{2001ApJ...547..792D} which extends to $|b| \leq 5^\circ$.
These observations have a spatial resolution of 8\,arcmin, $\sim$0.7\,km\,/\,s spectral resolution and $\sim0.3$\,K/channel sensitivity.

To study the radio continuum band, we made use of the Parkes-MIT-NRAO (PMN) survey, which was conducted at a frequency of 4.85\,GHz (\citet{1993AJ....105.1666G}; downloaded from ATNF\footnote{\url{ftp://ftp.atnf.csiro.au/pub/data/pmn//maps/PMN/}}).
The PMN survey has a resolution of 5\,arcmin, which was oversampled to 1\,arcmin per pixel.

For the infrared band, we use data from the Infrared Astronomical Satellite (IRIS) survey, observed at 12, 25, 60, and 100\,$\mu$m \citep{2005ApJS..157..302M}. 
The IRIS survey has a resolution of $\sim$\,1\,arcmin.

For the X-ray investigations, we use the MAXI Solid-state Slit Camera (SSC) all--sky maps (\citet{2014SPIE.9144E..1OM}; available from HEASARC\footnote{https://heasarc.gsfc.nasa.gov/docs/maxi/}).
The MAXI\,SSC energy bands overlap those of the ROSAT Position Sensitive Proportional Counter (PSPC) All-Sky Survey data (RASS) \citep{1993AdSpR..13l.391V}, which were used by \citet{2000ApJ...532..943C} in their initial assessment of the X-ray emission towards the Scutum Supershell. 
The MAXI\,SSC observations were taken during the solar minimum period of 2009 to 2011 \citep{2014SPIE.9144E..1OM}. 
In contrast, ROSAT\,PSPC  was in operation from 1990 to 1991 \citep{1995ApJ...454..643S}, including periods of high solar activity, which led to observations being affected by the Solar Wind Charge Exchange (SWCX) phenomenon \citep{1997ApJ...485..125S} (occurring when charged particles from the solar wind collide with neutral atoms in the Earth's atmosphere, creating an interfering source of X-ray photons). 
According to \citet{2016ApJ...829...83U}, the contamination from the SWCX in ROSAT\,PSPC data decreases from about 26\% to 6$\%$ across the 0.1 to 1.2\,keV range, suggesting a preferential focus on the energies approaching 1\,keV. 
The MAXI\,SSC observations feature an improved energy resolution of 0.1\,keV at 1\,keV \citep{2011PASJ...63..397T} compared to ROSAT\,PSPC's 0.4\,keV at 1\,keV\citep{1997ApJ...485..125S}.
MAXI SSC provides maps in the  0.7--1.0\,keV, 1.0--2.0\,keV, and 2.0--4.0\,keV energy bands, with a point spread function of 1.5$^\circ$ \citep{2009PASJ...61..999M}. 

Finally, to assess the TeV gamma-ray emission, we use the H.E.S.S. Galactic Plane Survey (HGPS; \citet{2018A&A...612A...1H}).
H.E.S.S.\ (High Energy Stereoscopic System) is an array of gamma-ray telescopes operating in an energy range from 30\,GeV up to 30\,TeV. 
The angular resolution of the HGPS is $\sim 0.1^\circ$, and the available flux maps provide the integral flux above 1\,TeV.
For the GeV gamma-ray emission, we made use of recent results (based on \textit{Fermi}-LAT observations) from \citet{2019MNRAS.485.1001A} and \citet{2021MNRAS.504.1840C}, who reported several GeV sources adjacent to the TeV-bright PWN HESS\,J1825$-$137.
\begin{figure*}
 	\centering 
    \includegraphics[width=0.95\textwidth]{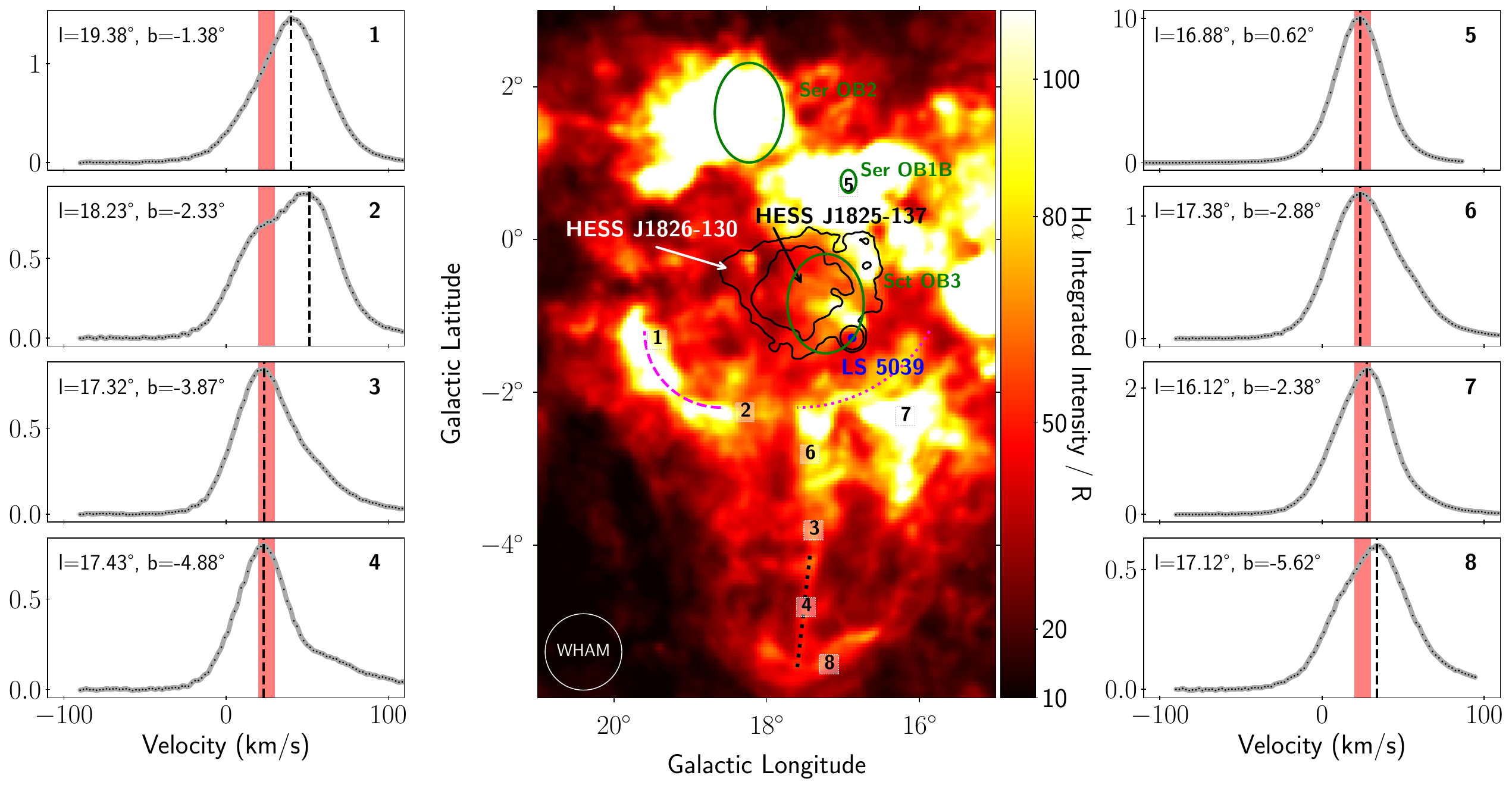}
    \caption{Wide view of optical H$\alpha$ emission. \textit{Middle}: Optical H$\alpha$ image from \citet{finkbeiner2003full} towards the Scutum Supershell. The TeV $\gamma$-ray significance contours obtained from HESS observations refer to the sources \PP{}, \PPP  and LS\,5039 (black contours of 5$\sigma$ and 10$\sigma$) \citep{2018A&A...612A...1H}. The OB associations \serob{}, Ser\,OB2, Sct\,OB3 \citep{1995Ast} are indicated as green ellipses. SNR\,G18.7--2.2 is indicated by a dashed magenta arc from \citet{stupar2008newly} and dotted from \citep{voisin2016ism}. The dotted black line highlights the spine feature. \textit{Left/right}: The eight panels show spectral profiles extracted from WHAM \citep{1998PASA...15...14R} for various pixels. 
    The red-shaded band (20--30\,km\,/\,s) indicates a kinematic distance of 2--3\,kpc. The black dots and gray lines represent the H$\alpha$ spectral intensities and B-spline interpolations. The vertical black dashed line indicates the velocity of the peak H$\alpha$ emission, $v_{\rm max}$.} 
    \label{V_l}  
\end{figure*}

\section{Results}

Here, we present a comprehensive examination of the H$\alpha$ and other multi-wavelength data towards the Scutum Supershell.

\subsection{H\texorpdfstring{$\alpha$}{}  Observations}
\label{sec:luminosity}

\autoref{fig:evo_h} compares the H$\alpha$ emission of the different datasets. 
A linear feature is clearly resolved in the SHS and Skywatcher images, and will be referred to as the `spine'.
Additionally, the southern end of the blowout clearly shows increased H$\alpha$ emission. 
We label this feature the `bow-shock' based on its apparent morphology. 
The linear spine feature appears to bisect the Scutum Supershell, and the bow-shock feature sits at the southern end of the spine and the southern boundary of the supershell. 

As a follow-up to Figure\,9 in \citet{2000ApJ...532..943C}, we show in \autoref{V_l} the H$\alpha$ image from \citet{finkbeiner2003full} along with example spectral profiles taken from single WHAM pixels.
The WHAM spatial FWHM of 1$^\circ$ is sufficient to spatially reveal the H$\alpha$ emission broadly along the outflow.
Based on this, we investigate the spectral information in the WHAM data to indicate the distance to the features and to compare to molecular and atomic gas associated with the supershell. 

\begin{figure}
    \centering
    \includegraphics[width=0.9\textwidth]{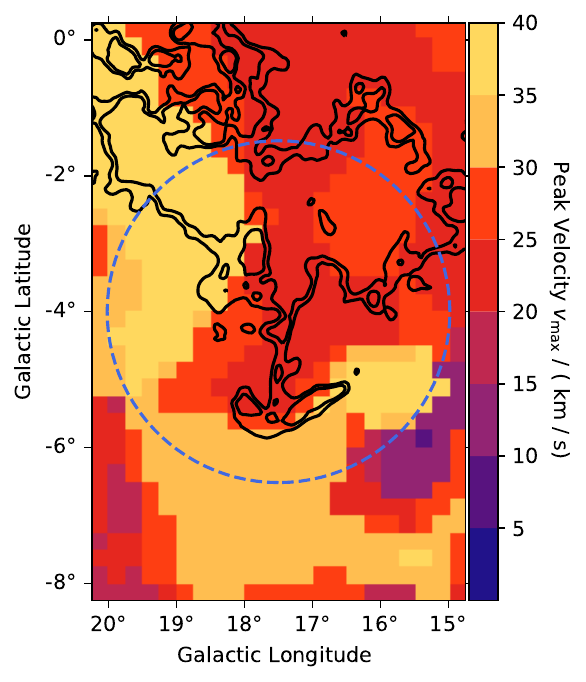} 
    \caption{Peak velocity distribution of WHAM H$\alpha$ emission spectra. The peak velocity $v_{\rm max}$ is determined from the first derivative of a B-spline interpolation of each pixel's spectrum. 
    The H$\alpha$ contours (black) indicate intensities of 40 and 50\,Rayleigh from the \citet{finkbeiner2003full} map, while the Scutum Supershell boundary is indicated by the dashed blue circle \citep{2000ApJ...532..943C}.}
    \label{fig:WHAM_max_inter}
\end{figure}

First, we infer approximate kinematic distances based on the Doppler-shifted velocity $v_{\rm max}$ at the intensity maximum of the spectral profile. 
To determine this velocity, we interpolate a function for each spectral profile using a B-spline\footnote{\texttt{splrep}, \texttt{splder}, \texttt{sproot} from \texttt{SciPy} \citep{2020SciPy-NMeth}}.
First derivatives of the B-spline were then used to estimate the velocity ($v_{\rm max}$) of the peak H$\alpha$ emission.
\autoref{V_l} shows spectral H$\alpha$ profiles along with their $v_{\rm max}$. 
We show H$\alpha$ spectra along the spine (spectra \#3, 4, 6), at the bow-shock to the south (spectrum \#8), the OB association \serob{}, and the \HII{} region G016.936+00.758 (spectrum \#5). 
The \HII{} region labelled S49 (see \autoref{fig:Star-pphotoion}) is associated with the Messier\,16 `Eagle Nebula' and overlaps with spectrum \#5.
The velocity range $20-30$\,km\,/\,s (shaded red bands in spectra of \autoref{V_l}) corresponds to a distance of about 2--3\,kpc using the Galactic rotation curve model by \citet{1993A&A...275...67B}.
The spectra \#1 and \#2 peak towards the 40--50\,km\,/\,s range, consistent with a $\sim4$\,kpc distance, however, the spectrum \#2 features a second peak around 20--30\,km\,/\,s.
The H$\alpha$ emission from these regions has been previously associated \citep{voisin2016ism,stupar2008newly} with the possible progenitor of SNR\,G18.7--2.2 (see \autoref{V_l}, magenta dashed arcs).
This SNR is linked to the TeV gamma-ray source \PP, and its distance is believed to be $\sim 4$\,kpc, based on the dispersion measure of the pulsar PSR\,J1826$-$1334 powering \PP\, and other interstellar gas studies by \citet{voisin2016ism}. 
The spectra from the other displayed regions (\#3 to \#8) peak within the 20--30\,km\,/\,s range. 
\autoref{fig:WHAM_max_inter} illustrates the peak velocities $v_{\rm max}$ of each WHAM pixel for an extended region around the Scutum Superhshell.
We can see that the H$\alpha$ emission associated with the spine and bow-shock features is broadly found in the 20--30\,km\,/\,s range.

Next, we investigate the different H$\alpha$ spectral components in more detail. 
We employ {\tt GaussPy+} \citep{2019A&A...628A..78R} to decompose each pixel's spectrum into multiple Gaussian components (see Appendix \ref{Gpy} for the results). 
The spectra are decomposed into up to five components; however, given WHAM's poor spectral resolution of $\Delta v=12$\,km\,/\,s {(this also defines the minimum width of any Gaussian feature) %which is considered in the fitting parameters of  {\tt GaussPy+})
}, and WHAM's preprocessing including a spectral oversampling%by a factor of 6
which leads to artifacts at large velocities {$\gtrsim$80\,km\,/\,s} 
, we focused on the main (significant) components.
We find that all spectra are decomposed into one or two main components.
The spectra of the pixels encompassing the spine and bow-shock features have typically one main component peaking in the $20-30$\,km\,/\,s range, supporting our results from the analysis of the peak velocity and our estimate of associating the spine with a distance of $2-3$\,kpc.

From a visual inspection of the high-resolution H$\alpha$ images, the spine is visible between Galactic latitudes of $-5.5^{\circ}$ and $-3.5^{\circ}$ for about 2$^\circ$ in length, which corresponds to a physical length of $l \sim 70$\,pc, assuming a $2.5$\,kpc distance. 
Unfortunately, the bright H$\alpha$ emission further north associated with \HII{} regions and other potential energetic objects (e.g.\ SNR shocks) complicates our determination if the spine extends further.

Some further insight into the origin of the H$\alpha$ emission can come from the corresponding intrinsic luminosity. 
To extract the luminosity of the spine and of the bow--shock region, we used data from the H$\alpha$ surveys WHAM, and \citet{finkbeiner2003full}. %SHASSA and SHS individually. 
Background emission was estimated from several regions, as listed in \autoref{tab:Bakcground_ha}, and was subtracted to estimate the luminosity above the local background. 
The luminosity $L$ is estimated as $L = 4\pi d^2 I$ for intensity $I$ and distance $d$. 
We find H$\alpha$ luminosities of 
$$L_{{\rm H}\alpha} (2.5\,\rm kpc / d)^2  \sim 10^{36} \,erg\,\rm /s\ ;$$
see \autoref{tab:L} for results of individual surveys and regions.
For comparison, \autoref{tab:L_other_source} provides H$\alpha$ luminosities of other astrophysical objects that also exhibit H$\alpha$ spine and bow-shock features, ranging from protostellar Herbig--Haro objects to X-ray binaries and active galaxies.

\begin{table}  
    \centering
    \caption{H$\alpha$ luminosities $L_{{\rm H}\alpha}$ for the spine and bow-shock features. \autoref{tab:Bakcground_ha} lists the signal and background regions and their corresponding extracted luminosities.}
    \begin{threeparttable}
        \begin{tabular}{lc}
        \toprule
        Region & $L_{{\rm H}\alpha}(2.5\,{\rm kpc} \, / \,d)^2 $  \\
         & (erg\,/\,s)    \\ 
        \midrule
        H$\alpha$ bow-shock (Finkbeiner)  &  2.07$\times$10$^{36}$ \\
        H$\alpha$ bow-shock (WHAM)  &  1.31$\times$10$^{36}$ \\
        H$\alpha$ spine (Finkbeiner)
        &  1.23$\times$10$^{36}$ \\ 
        H$\alpha$ spine (WHAM) 
        & 0.78$\times$10$^{36}$ \\
        \bottomrule
        \end{tabular}
    \end{threeparttable}
    \label{tab:L}
\end{table}

\subsection{[S II] Observations}

In addition to the H$\alpha$ emission, other lines such as [S\,II], [N\,II] and [O\,III] help to diagnose the role of shocks and photo-ionisation as the excitation mechanism for the gas.
For example, \citet{1985ApJ...292...29F} and \citet{1977ApJS...33..437D} demonstrate that [S\,II] emission is an important indication of shocked gas from photo-ionised regions.
This is due to the high electron temperature in the ionised sulfur S$^+$ zone of the shock, where non-equilibrium recombination occurs.
The production of [S\,II] in a shock model requires a high shock velocity or gas pressure along with a high magnetic-field pressure \citep{1979ApJS...39....1R}.

\begin{figure}
    \centering
    \includegraphics[width=\textwidth]{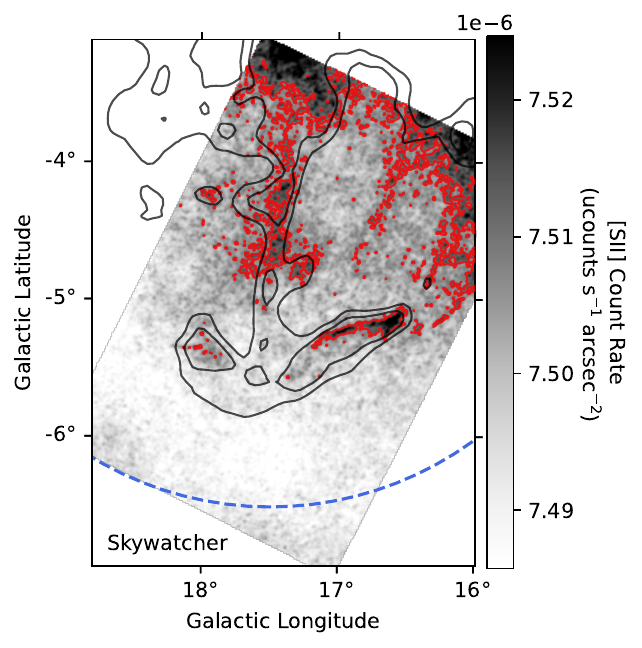}
    \caption{Skywatcher [S II] (uncalibrated) count rate towards the Scutum Supershell (dashed blue circle). The red contours indicate observations at 5\,$\sigma$ significance level.
    H$\alpha$ emission from the composite map at 40 and 50\,R \citep{finkbeiner2003full} is illustrated as black contours.}
    \label{fig:sii_image}
\end{figure}

Our work is the first to map the [S\,II] optical emission at arcsecond resolution over this region, enabling a comparison with the H$\alpha$ emission.
\autoref{fig:sii_image} clearly shows [S\,II] emission in the bow--shock region of the H$\alpha$ emission, peaking at the right side of the bow-shock. 
The emission line ratio [S\,II]/H$\alpha$ can be used to infer the physical conditions of the ionised gas \citep{1979ApJS...39....1R}, such as the potential for shock-excitation or photo-ionisation. 
However, since our [S\,II] observations are not flux-calibrated and our [S\,II] filter included both the [S\,II] 671.6\,nm{} and 673.1\,nm{} lines, we cannot determine this ratio.

\subsection{Radio Continuum Observation}

\begin{figure}
    \centering
    \includegraphics[width=\textwidth]{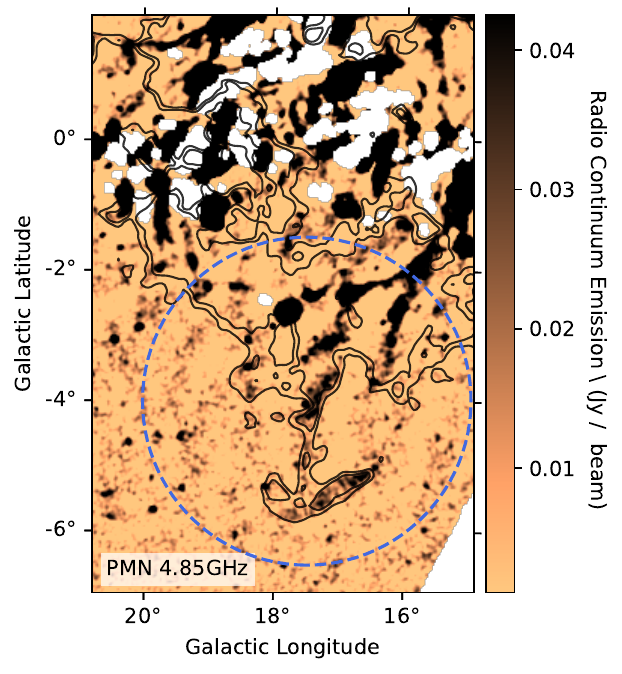}
    \caption{PMN 4.85\,GHz radio continuum emission towards the Scutum Supershell (dashed blue circle). H$\alpha$ emission \citep{finkbeiner2003full} at 40 and 50\,Rayleigh is illustrated as black contours. Saturated pixels (>40\,Jy; \citet{1993AJ....105.1666G}) are shown in white, as well as pixels outside the observed field.}
    \label{Fig:PMN_radio}
\end{figure}

\autoref{Fig:PMN_radio} shows the PMN 4.85\,GHz survey \citep{1993AJ....105.1666G} radio continuum emission. 
Interestingly, we find that the emission overlaps the brightest H$\alpha$ and [S\,II] emission at the right side of the bow-shock.

\subsection{Infrared Observations}\label{sec:IRIS}

\begin{figure*}
    \centering
    \includegraphics[width=\textwidth]{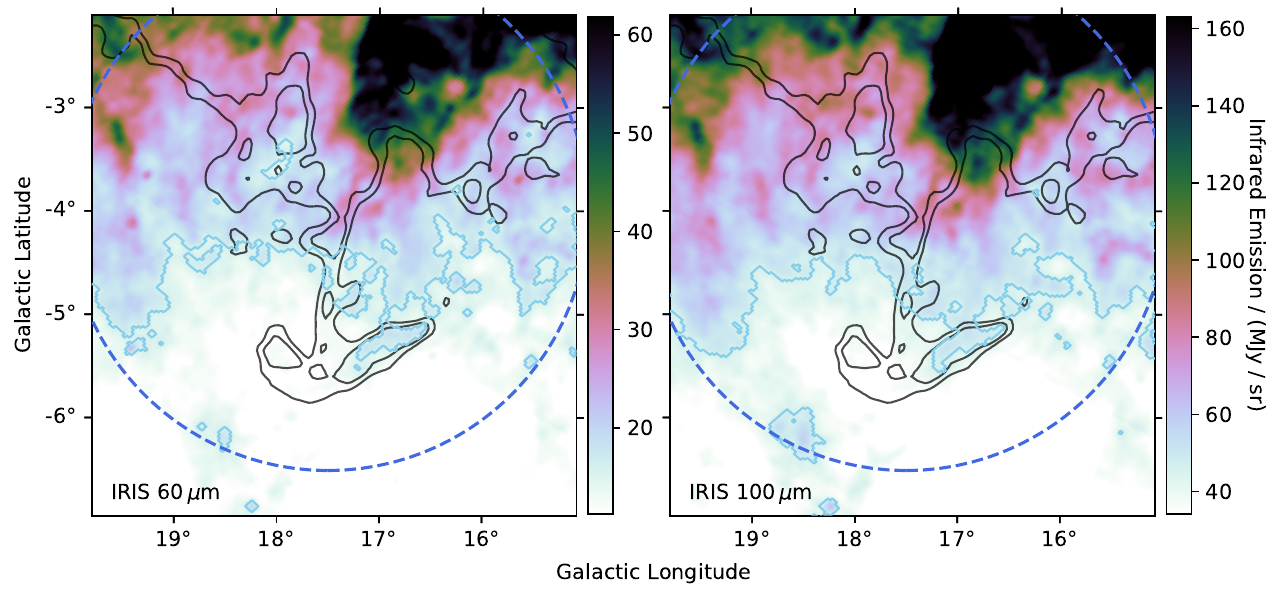}
    \caption{IRIS infrared emission \citep{2005ApJS..157..302M} in the 60\,$\mu$m (left) and 100\,$\mu$m (right) bands towards the H$\alpha$ emission (black contours). Lightblue contours represent emission with a significance of 5$\sigma$.}
    \label{Fig:IRIS_bands}
\end{figure*}

The IRIS \citep{2005ApJS..157..302M} 12 and 25\,$\mu$m images did not show any significant emission towards the H$\alpha$ features. 
However, we observe emission (see \autoref{Fig:IRIS_bands}) in the 60 and 100\,$\mu$m observations towards the bow-shock region overlapping the brightest H$\alpha$ and [S\,II] features and the 4.85\,GHz radio continuum emission.
The estimated fluxes for the region with significant infrared emission (right side of the bow-shock) are $\sim 2.4 \times10^{-5}$\,MJy (60\,$\mu$m) and $\sim 6.1\times10^{-5}$\,MJy (100\,$\mu$m). 
For these flux {density} estimates, we subtracted a background estimate using the regions listed in \autoref{tab:IRIS_bk}. 

\subsection{Atomic and Molecular Hydrogen Observations}

\begin{figure*}
    \centering 
    \includegraphics[width=1\textwidth]{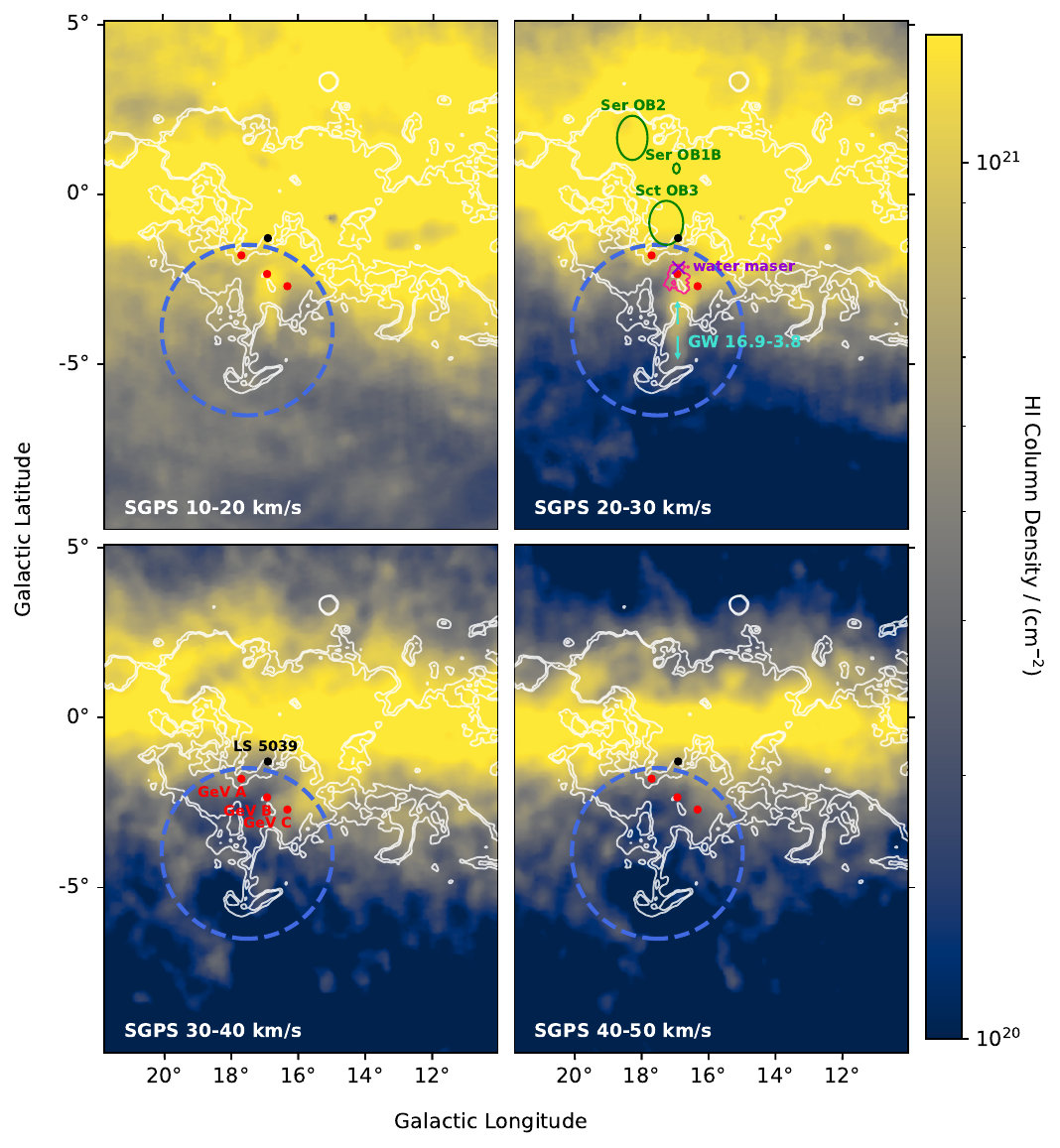}
	\caption{SGPS \HI{} column density maps \citep{2005ApJS..158..178M} for various velocity ranges. H$\alpha$ emission is indicated as white contours at 40 and 50\,Rayleigh \citep{finkbeiner2003full}. The Scutum Supershell is indicated by the blue dashed circle. The {\em Fermi}-LAT GeV sources (GeV\,A,~B,~and~C) are marked with red dots \citep{2019MNRAS.485.1001A}, and the binary LS\,5039 with a black dot. The pink contours in the 20--30\,km km\,/\,s map show the molecular hydrogen Nanten\,CO(1--0) emission (from \citet{2004ASPC..317...59M} for the (15--30)\,km\,/\,s range). The same velocity range also encompasses the water maser G016.8689$-$02.1552 (purple cross) \citep{2011MNRAS.418.1689U}. The \HI{} Galactic worm, GW\,16.9$-$3.8, and its orientation are indicated in turquoise, and the green dashed ellipses mark the OB associations Ser\,OB2, \serob{} and Sct\,OB3. }
	\label{SGPSHI}
\end{figure*}

As shown by \cite{2000ApJ...532..943C}, the \HI{} emission reveals the atomic gas boundary associated with the Scutum Supershell.
\autoref{SGPSHI} shows \HI{} column density distributions in the range between 10 and 50\,km\,/\,s, each integrated over 10\,km\,/\,s. 
For converting the \HI{} brightness temperature to column density, we used the conversion factor from \citet{1990ARA&A..28..215D}, which assumes optically thin \HI{} emission.

In the 20$-$30\,km\,/\,s range, we noticed that the Galactic worm GW\,16.9$-$3.8, found by \citet{1992ApJ...390..108K} in \HI{} emission, is adjacent to the H$\alpha$ spine.
With an estimated distance of 2.5\,kpc, the worm has a length of 50 to 180\,pc.
\citet{1992ApJ...390..108K} and \citet{2000ApJ...532..943C} suggested that this worm is at the foreground edge of the Scutum Supershell, and may be driven by stellar winds or supernovae from the \serob{} association \citep{1992ApJ...390..108K} (also indicated in \autoref{SGPSHI}). 

The \HI{} maps in the 10 to 30\,km\,/\,s range also reveal strong emission towards the GeV\,B gamma-ray peak. 
The three GeV gamma-ray peaks, labelled GeV\,A, GeV\,B and GeV\,C, were discovered by \citet{2019MNRAS.485.1001A} in their analysis of \textit{Fermi}-LAT GeV data. 
They noted that the three peaks are found several degrees south of the PWN HESS\,J1825$-$137 and \citet{2021MNRAS.504.1840C} considered their origin due to particles escaping the PWN.

\citet{2001PASJ...53.1003M} showed several Galactic supershells emitting $^{12}$CO.
\autoref{fig:COmap} shows the $^{12}${CO(1--0)} distributions from the CfA-Chile CO survey. 
These maps cover an expanded region (down to $b=-5^\circ$) compared to the Nanten $^{12}$CO(1--0) maps shown by \citet{2021MNRAS.504.1840C}.
We used the X--factor {$X_{\rm CO}=1.5\,\times 10^{20}$\,cm$^{-2}/$\,K\,km\,s$^{-1}$}
from \citet{2004A&A...422L..47S} to convert to a molecular hydrogen column density. % $N_{\rm H_2}$.
We see a molecular cloud feature in the 15 to 30\,km\,/\,s range towards the GeV\,B gamma-ray peak, as pointed out by \citet{2019MNRAS.485.1001A} and also by \citet{2021MNRAS.504.1840C}, who used higher-resolution Nanten $^{12}$CO(1--0) data (we show the Nanten contours for this molecular feature in \autoref{SGPSHI}). 
Interestingly in the same velocity range, the water maser G016.8689$-$02.1552
is found towards this molecular cloud, indicating it is active in star formation \citep{2011MNRAS.418.1689U,2016ARep...60..438K}. 
The maser is likely associated with L379\,IRS1, a massive star-formation region located at a distance of 2.5\,kpc \citep{2016ARep...60..438K}.

\subsection{X--Ray Observations}\label{sec:x-ray}
 
 \cite{2000ApJ...532..943C} used the ROSAT\,PSPC all-sky survey in the 0.1--0.28, 0.4--1.2, and 0.7--2.0\,keV energy ranges to investigate the X-ray emission. 
They suggested that the X-ray emission may originate from hot gas driven by the stellar winds and SNRs that power the Scutum Supershell. 
\autoref{fig:xray-wide}  shows wide-scale views of the MAXI\,SSC 0.7--1.0\,keV and ROSAT\,PSPC 0.4--1.2\,keV X-ray photon fluxes. 
We can see that both datasets reveal X-ray emission peaking locally towards the Scutum Supershell and extending $\sim 5^\circ$ south of the H$\alpha$ feature (black contours).

\begin{figure}
    \centering 
	\includegraphics[width=\textwidth]{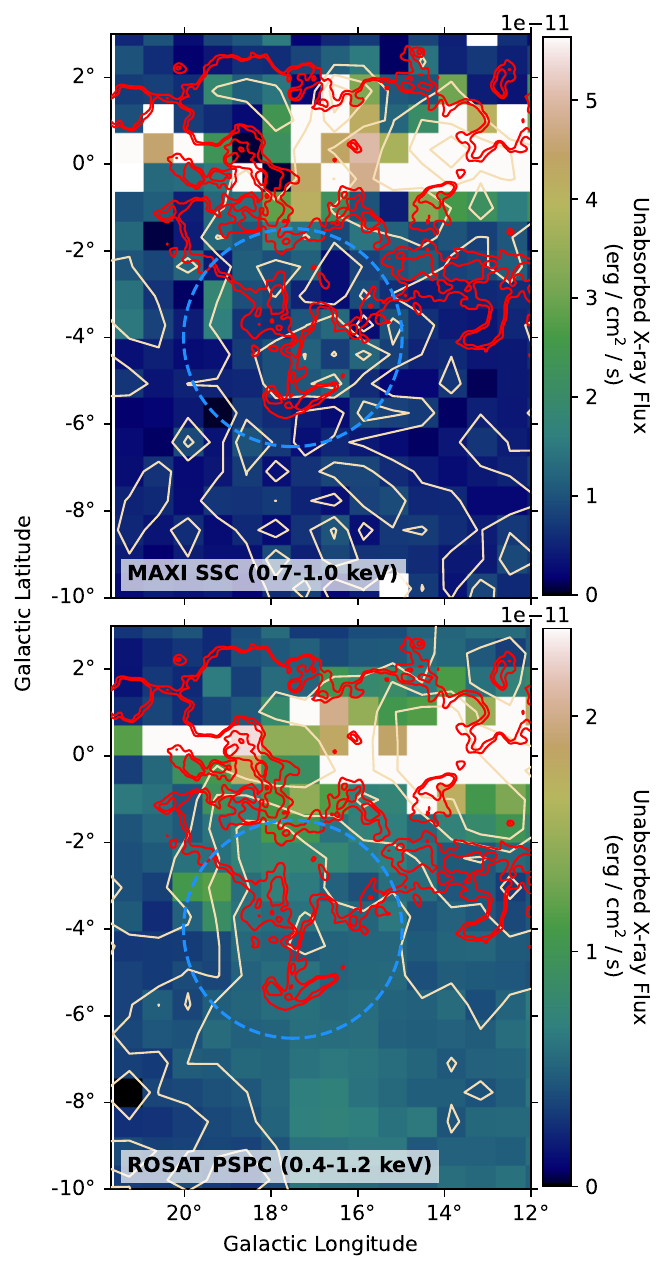}
	\caption{Soft-band X-ray unabsorbed flux maps for both ROSAT\,PSPC and MAXI\,SSC.
 The H$\alpha$ contours (red) are shown at the 40\,R and 50\,R levels. X-ray contours (white) are shown at the 5, 10, and 15\,$\sigma$ level. 
    The Scutum Supershell boundary is illustrated as a blue dashed circle \citep{2000ApJ...532..943C}. 
    }
	\label{fig:x-r}
\end{figure}

We broadly replicated the analysis performed by \citet{2000ApJ...532..943C} in their analysis of the ROSAT PSPC data. 
We first converted the MAXI\,SSC and ROSAT PSPC photon flux maps to an absorbed (observed) energy flux via the online Portable, Interactive, Multi-Mission Simulator (PIMMS) tool \citep{1993Legac...3...21M}. 
For this step, we assumed a black-body thermal model (with temperature $T$) for the X-ray emission, and used an upper-temperature limit of $T=10^7$\,K, similar to \citet{2000ApJ...532..943C}, which is suitable for the production of diffuse soft X-ray emission from thermal plasma processes \citep{2004ASSL..309..103S}. 
The PIMMS tool estimates the unabsorbed energy flux by accounting for photoelectric absorption from the total column density ($\textrm{n}_{\mathrm{H}}$) of the foreground gas from the \HI{} and CO observations (\citet{2005ApJS..158..178M}, \citet{2001ApJ...547..792D}) integrated over the velocity range from 0 to 30\,km\,/\,s. 
The above-mentioned process was repeated on a pixel-by-pixel basis using a 40\,arcmin$\times$40\,arcmin binning (a factor 4 larger than the original MAXI\,SSC pixels), and the \HI{} and CO data were regridded to match the X-ray data.
The unabsorbed fluxes in the soft band are directly compared between MAXI and ROSAT in \autoref{fig:x-r} in relation to the H$\alpha$ emission.
\autoref{fig:x-soft} and \autoref{fig:x-hard} shows the photon fluxes and the absorbed and unabsorbed energy fluxes for both ROSAT\,PSPC and MAXI\,SSC observations. 

Overall, the MAXI\,SSC images reveal similar large-scale X-ray structures overlapping the Scutum Supershell to those highlighted by \citet{2000ApJ...532..943C} using ROSAT data. 
The X-ray emission appears inside the Scutum Supershell boundary, encompassing the H$\alpha$ features.

To estimate the X-ray unabsorbed luminosity within the Scutum Supershell, we selected a region identical to \citep{2000ApJ...532..943C}, and used PIMMS to convert the energy flux to a luminosity, after subtracting the X-ray emission from the background regions defined in \autoref{tab:maxi_bkgd} and using the column density from the \HI{} and CO data averaged over the region, and a temperature of  $T=10^7$\,K for the black-body model.
We obtained a luminosity $L_{\textnormal{\scriptsize X-ray}}$ from the unabsorbed X-ray emission of 
$$L_{\textnormal{\scriptsize  X-ray, ROSAT}} = 3.9 \times 10^{36}(d/2.5\,{\rm kpc})^2\,{\rm erg}\,{\rm / s}$$
for ROSAT PSPC (0.4--1.2\,keV) and 
$$L_{\textnormal{\scriptsize X-ray, MAXI}} = 4.6 \times 10^{36}(d/2.5\,{\rm kpc})^2\,{\rm erg}\,{\rm / s}$$
for MAXI\,SSC (0.7--1.0\,keV). 
For completeness, \autoref{tab:X-ray_pimf} lists the results for different temperatures. 
We conclude that our PIMMS-derived luminosity is consistent (within 4\%) to those obtained by \citet{2000ApJ...532..943C}.

\section{Discussion}
\label{discuss}

We will now review the multi-wavelength emissions towards the Scutum Supershell and discuss the potential origins of the H$\alpha$ spine and bow-shock features.

The X-ray maps from MAXI\,SSC exhibit comparable morphology and fluxes to the ROSAT\,PCPS observations (compare \autoref{tab:X-ray_pimf} and \citet{2000ApJ...532..943C}).
Our analysis with PIMMS indicates that the soft-band luminosity of both the MAXI SSC and ROSAT\,PSPC X-ray emissions is  approximately $10^{36}(d/2.5\,{\rm kpc})^2$erg\,/\,s.
Morphologically, the thermal soft-band X-ray emission (see \autoref{fig:x-r}), which is present in the Scutum Supershell and extends further south, indicates the presence of hot gas (as also noted by \citet{2000ApJ...532..943C}).

\begin{figure}[!htb]
\centering
\includegraphics[width=\textwidth]{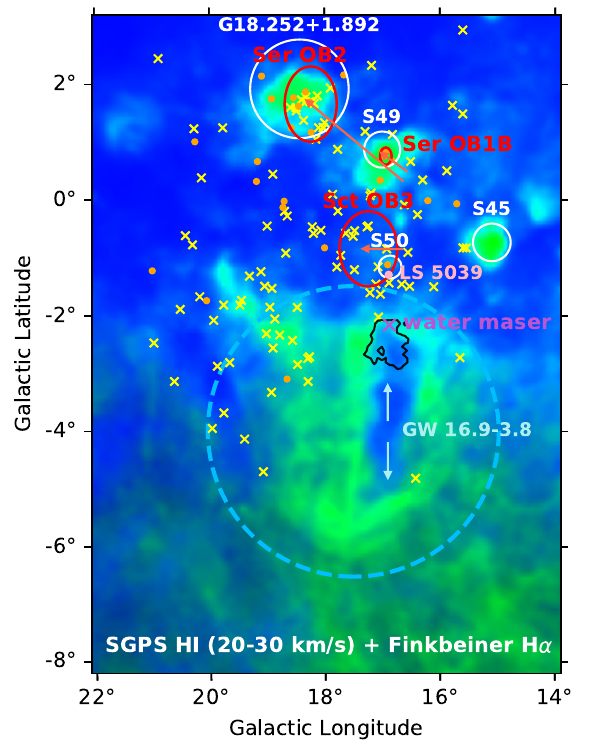}
 \caption{ 
        Composite of H$\alpha$ emission (green) \citep{finkbeiner2003full} and \HI{} emission (blue) \citep{2005ApJS..158..178M} integrated over the 20 to 30\, km\,/\,s range. The Galactic worm GW\,16.9$-$3.8 is the prominent \HI{} vertical feature seen surrounded by the H$\alpha$ emission. The CO(1--0) emission \citep{2004ASPC..317...59M} for the 15--30 km\,/\,s range (black contours) shows the molecular hydrogen encompassing the water maser G016.8689$-$02.1552 (purple cross) \citep{2011MNRAS.418.1689U}.
         The proper motions (red arrows) of the star clusters linked to the OB associations (Ser\,OB2, \serob{}, Sct\,OB3; red ellipses) indicate their location 1\,Myr ago (\autoref{tab:ob_assoc}).  
         White circles mark \HII{} regions \citep{2014ApJS..212....1A} with velocities 15--30 km\,/\,s and radius of >500\,arcsec. Stars confirmed as O-type (orange dots) and those catalogued as OB (uncertain O or B; yellow crosses) \citep{2003AJ....125.2531R} were considered in the photo-ionisation calculations.
         LS\,5039 (pink dot) is a high-mass X-ray binary.} 
	\label{fig:Star-pphotoion}
\end{figure}

The SGPS \HI{} observations reveal a broadly similar morphology to those observed by \cite{2000ApJ...532..943C} using earlier NRAO (VLA) \HI{} data. 
However, closer inspection of the SGPS \HI{} image in the 20--30\, km\,/\,s range, compared to the H$\alpha$ emission reveals a spatial anti-alignment of the spine with the Galactic worm GW\,16.9$-$3.8 seen in \HI{} (see \autoref{fig:Star-pphotoion}). 
GW\,16.9$-$3.8 was first discussed by \citet{2000ApJ...532..943C} and \citet{1992ApJ...390..108K} as potentially having some \blue{relation to} the Scutum Supershell. 
The Galactic worm overlaps a void in the H$\alpha$ emission, and %\red{thus appears to act as a shield for any ionisation that may lead to such emission.}
 as a result, it is either shielding the region behind it from ionising radiation. Thereby preventing H$\alpha$ emission, or the H$\alpha$ emission behind absorbed by \HI{} and dust in front, making it invisible to our observations. 
The spine feature may therefore represent the ionised surface of the Galactic worm GW\,16.9$-$3.8.
The northern edge of the worm hosts a molecular gas component based on $^{12}$CO emission in an overlapping velocity range (15--30\,km\,/\,s) from Nanten observations, and a water maser G016.8689$-$02.1552 \citep{2011MNRAS.418.1689U,1996ApJ...466..191H}, suggesting the presence of active star formation. 

\autoref{fig:Star-pphotoion} also shows the location of O and OB stars in the distance range of 1 to 3\,kpc, along with the OB associations  \serob{}, Ser\,OB2, and  Sct\,OB3. 
These three OB associations have recent distance estimates of $1.3-1.7$\,kpc (see \autoref{tab:photoionisation_OB}), based on Gaia DR3 measurements.
The star clusters associated with the OB associations all have proper motions in the range 0.2 to 7\,mas/yr (\autoref{tab:ob_assoc}), which means they have travelled $\sim 1^\circ$ or more over their life times. 
It should be noted that \serob{} has been linked to GW\,16.9-3.8 and two other Galactic worms GW\,14.9-1.6 and GW\,19.5-6.4 \citep{1992ApJ...390..108K}. 
GW\,16.9-3.8 is centered at a velocity of 25\,km\,/\,s, giving a near distance solution of 2.5\,kpc using the Galactic rotation model of \cite{1993A&A...275...67B}.
The somewhat lower GAIA-derived distance for \serob{} might suggest the local velocity of  GW\,14.9-1.6 is influenced by the expansion of the Scutum Supershell by 10\,km\,/\,s or so.

Interestingly, the H$\alpha$ bow-shock feature corresponds to a peak in the [S\,II] emission towards its rightmost edges according to our Skywatcher observations (see \autoref{fig:sii_image}). 
At the rightmost edge there is also overlapping 4.85\,GHz radio continuum emission (see \autoref{Fig:PMN_radio}), and overlapping infrared emission at 60$\mu$m and 100$\mu$m (see \autoref{Fig:IRIS_bands}). 
The H$\alpha$ spine has some overlapping [S\,II] emission but at a weaker level than for the bow-shock region. No other obvious overlapping radio or infrared emission is seen.

Overall, given the presence of many potential stellar and non-stellar sources of ionisation in the region, we will consider both photo-ionisation and shock-induced excitation processes for both the spine and the bow-shock H$\alpha$ emission features.

\begin{table*}[ht]
    \caption{Expected H$\alpha$ luminosity $L_{\rm exp}$ created by Lyman continuum photons from the OB associations Ser\,OB1B, Ser\,OB2 and Sct\,OB3 as sources of photo-ionisation. The luminosities are estimated for the bow-shock region (dotted rectangle in \autoref{fig:HII_region}). The surface brightness $S$ and corresponding surface area $A$ are used to estimate the absorption photon rate $Q_{\rm env}$. The number of stars (\# Stars) in each OB association is taken from \citet{Stoop2023, 2018MNRAS.473..849D}, the number of O-type stars from \citet{1995Ast, Stoop2023}, and the remaining stars are considered B-type stars. These numbers are used together with photon rates for individual stars of a specific spectral type from \citet{1996ApJ...460..914V} to calculate the photon rate $Q_{\star}$.
    }
    \centering
    \begin{tabular}{lccccccc|cc|cc}
    \toprule
    Name & Dist. & \multicolumn{2}{c}{\#\,Stars} & $Q_{\star}$ & $S$ & $A$ & $Q_{\rm env}$ & $r_{d=2.5 \rm kpc}$ & ${L_{{\rm exp, }d=2.5 {\rm kpc}}}$ & $r_{d=1.5 \rm kpc}$ & ${L_{{\rm exp, }d=1.5 {\rm kpc}}}$ \\
       & (kpc) & O-type & B-type & ($10^{50}$\,/\,s) & (Rayl.) & (pc$^2$) & ($10^{48}$\,/\,s) & (pc) & (erg\,/\,s) & (pc)  & (erg\,/\,s) \\
    \midrule
    Ser OB1B & 1.71$^{\tiny a}$ & 19$^{\tiny a}$ & 118$^{\tiny a}$ & 11.3 & 1702 & 42 & 2.4 & 826 & $3.4 \times 10^{36}$ & 270 & $8.6 \times 10^{36}$ \\
    Ser OB2 & 1.60$^{\tiny b}$ & 8$^{\tiny c}$ & 252$^{\tiny d}$ & 5.4 & 233 & 719 & 5.7 & 935 & $1.4 \times 10^{36}$ & 219 & $5.6 \times 10^{36}$ \\
    Sct OB3 & 1.33$^{\tiny b}$ & 2$^{\tiny c}$ & 143$^{\tiny d}$ & 2.0 & 65 & 550 & 1.2 & 1179 & $0.4 \times 10^{36}$ & 205 & $2.3 \times 10^{36}$ \\
    \bottomrule
    \end{tabular}
    \begin{tablenotes}
            \item[$a$] \citet{Stoop2023}
            \item[$b$] \citet{2020NewAR..9001549W}
            \item[$c$] \citet{1995Ast}
            \item[$d$] \citet{2018MNRAS.473..849D}: Total number of stars of any spectral type
    \end{tablenotes}
  \label{tab:photoionisation_OB}
\end{table*}

\subsection{Photo-ionisation Excitation}\label{sec:photoion}

Here, we will consider the possibility of Lyman continuum (LyC) photons as a source of ionisation for the H$\alpha$ emission. 
The LyC photons could arise from O and B stars from within the OB associations or from additional O and B stars from a wider region. 

We employed the method from \citet{1997ApJ...474L..31D}, which calculates the expected luminosity of H$\alpha$ emission created by LyC photons some distance away.  
\blue{Some of the LyC photons will be absorbed via { photoionisation.} {Subsequent a} recombination to produce H$\alpha$ emission in surrounding compact \HII{} regions or in broader regions farther out.} 
We can compare this expectation to the observed H$\alpha$ luminosity from the bow-shock region which is 

$$L_{\mathrm{obs, H}\alpha} \sim 2 \times 10^{36} \, (d / 2.5\,{\rm kpc})^2\,\rm erg\,\rm / s \ .$$ 

By rearranging Eq.\,1 of \citet{1997ApJ...474L..31D} and transforming to luminosities, the expected luminosity $L_{\mathrm{exp, H}\alpha}$ of the H$\alpha$ emission from a region of distance d from Earth, and with solid angle $\Omega$, thickness $s$  and distance $r$ from a LyC photon source is calculated via:
{\large
\begin{eqnarray}
      L_{\mathrm{exp, H}\alpha} =& \frac{4 \, \pi \, {\rm d}^2 \, \Omega \, \left(Q_{\star} / 4\pi \, - \, Q_{\rm env} / 4\pi \right)} {r^2 \, \alpha _B (T) \, \left(1 + 2r / s\right)^{-0.5} \, \left(T / 10^4\,{\rm K}\right)^{0.9}} \nonumber \\
      &\times \ \frac{\mathrm{erg} \, / \, \mathrm{s} \, / \,  \mathrm{cm}^2 \, / \, \mathrm{arcsec}^2}{4.858 \times 10^{17} \, \mathrm{pc} \, / \, \mathrm{cm}^6} \ .
      \label{eq:photi1}
\end{eqnarray}
}
${Q_{\star}}$ is the LyC photon rate for the stellar source in question, either a grouping of stars or an individual star. 
We obtained ${Q_{\star}}$ by summing over individual photon rates, which we obtained from \citet{1996ApJ...460..914V} for individual O and B stars dependent on their spectral type. 
$Q_{\rm env}$ is the effective absorption LyC photon rate required to match the observed H$\alpha$ surface brightness $S$ assumed to be due to recombination from the `environment' (env), a region surrounding the photon source, or, a region in between the photon source and the region of interest.
Here, we calculate this rate based on the surface area $A$ and H$\alpha$ surface brightness $S$ of the surrounding region. 
\begin{equation}
    Q_{{\rm env}} = 2.75 \, \frac{{\rm pc}}{{\rm cm}^6} \, \frac{S}{{\rm Rayleigh}} \, \left(\frac{T}{10^4\,{\rm K}}\right)^{0.9} A \, \alpha_{\rm A}(T) \ .
\end{equation}
As per \citet{1997ApJ...474L..31D}, we assume an edge-brightened shell geometry for the region of interest (the bow-shock).   
The recombination rates $\alpha_{\rm A}=4.18\times 10^{-13}$ and  $\alpha_{\rm B}=2.59\times 10^{-13}$ cm$^3$\,/\,s are taken from  \citet{1989agna.book.....O} for $T=10^4$\,K.

We first calculated the expected H$\alpha$ luminosities for the three OB associations as the sources of LyC photons, and the bow-shock region with thickness of $s = d \cdot \tan(0.8^\circ)$ (the extension of the rectangular bow-shock region in North-South direction), centred at $l=17.3^\circ$ and $b=-5.5^\circ$ (this region is shown in \autoref{fig:HII_region}), and assuming a distance of $d=2.5$\,kpc. 
\serob{} and the \HII{} region Messier 16 (S49) are related to the star cluster NGC\,6611 \citep{2019MNRAS.484.4529B} which contains 51 stars \citep{2018MNRAS.473..849D}. 
Recent work by \citet{Stoop2023} states that NGC\,6611 hosts 19 O-type stars (with different provided spectral subclasses) and 137 stars in total at a distance of 1.706\,kpc, based on \textit{Gaia} EDR3. 
Here, we will assume the remaining 118 stars are early B-type stars. 
For Ser\,OB2, \citet{2000AJ....120.2594F} suggested a relationship to the \HII{} region Sh2--54 (G018.252+1.892), and the star cluster NGC\,6604. 
Ser\,OB2 and its star cluster NGC\,6604 contain 260 stars \citep{2018MNRAS.473..849D}, eight of which are O-type stars \citep{1995Ast}. 
The remaining 252\,stars will be assumed to be B-type stars. 
Sct\,OB3 contains 145 stars \citep{2018MNRAS.473..849D}, two of which are O-type stars as noted by \citet{1995Ast}. 
The Wolf--Rayet star WR\,115 is another member of Sct\,OB3.
We will assume the remaining 143 stars are B-type.
The resulting ${Q_{\star}}$ for the above assumptions are listed in \autoref{tab:photoionisation_OB}.
The number of stars belonging to the OB associations and the corresponding spectral subtypes, together with the assumption that the remaining stars are B-type stars introduce some uncertainty. 
If we only include the O stars in the calculation of the photon rate ${Q_{\star}}$, we obtain $2.9 \times 10^{50}$\,/\,s for Ser\,OB2, $0.6 \times 10^{50}$\,/\,s for Sct\,OB3, and $10.1 \times 10^{50}$\,/\,s for Ser\,OB1B.
To estimate $Q_{\rm env}$, we extracted the H$\alpha$ surface brightness from the elliptical extensions of the OB associations (see \autoref{fig:Star-pphotoion} and \autoref{tab:ob_assoc}). 
This value depends on the choice of the extraction region and increased to $\sim 8 \times 10^{48}$\,/\,s for Ser\,OB2 when increasing the extraction region to encompass the bright H$\alpha$ emission in that region, varied between $(1.0-1.5) \times 10^{48}$\,/\,s for Sct\,OB3 when varying the ellipticity of the extraction region, and increased to $\sim (10-13) \times 10^{48}$\,/\,s for Ser OB1B when increasing the region to encompass the bright H$\alpha$ emission. 
The results in \autoref{tab:photoionisation_OB} show that the expected H$\alpha$ luminosity from each of the OB associations $L_{\mathrm{exp, H}\alpha}$ matches that of the observed luminosity of $\sim 2 \times 10^{36}(d/2.5\,{\rm kpc})^2$\,erg\,/\,s for the bow-shock region. 
Using the decreased ${Q_{\star}}$ and the increased $Q_{\rm env}$, we obtain luminosities ${L_{{\rm exp, }d=2.5 {\rm kpc}}}$ of $0.7 \times 10^{36}$\,erg\,/\,s for Ser\,OB2, $0.1 \times 10^{36}$\,erg\,/\,s for Sct\,OB3, and $3.0 \times 10^{36}$\,erg\,/\,s for Ser\,OB1B.
If instead we assume the H$\alpha$ features are at a distance $d=1.5$\,kpc (and adjust $r$ and $s$ accordingly) to match the GAIA-derived distances of the OB associations, the expected luminosity ${L_{\rm exp,1.5}}$ increase.
We conclude that the OB associations could be photo-ionisation sources of the observed H$\alpha$ features. 

In the second application of \autoref{eq:photi1}, we assume the LyC photons are produced from massive stars specifically catalogued across the field extending beyond the three OB associations. 
We used the \cite{2003AJ....125.2531R} catalogue and \texttt{Simbad} \citep{2000A&AS..143....9W} to identify O and OB stars within 1 to 3\,kpc distance from Earth, and Galactic coordinates $15.5^\circ \leq l \leq 21^\circ$ and  $-5^\circ \leq b \leq 5^\circ$.  
We found a total of 33 O stars with identified spectral sub-types, and an additional 103 stars listed as `OB' without spectral sub-types. 
The O and OB stars  in the field of the Scutum Supershell are shown in \autoref{fig:Star-pphotoion} with the full lists given in \autoref{tab:star_photo_fesc} and \autoref{tab:star_photo_fesc2}. 
Here, we will assume a photon rate $Q_{\star}$ of $10^{49}$\,/\,s for stars labelled `OB', corresponding to the rate of stars with a spectral type O9.
For the absorption rate $Q_{\rm env}$, we choose circular regions of radius $0.2^\circ$ around each star to extract the surface brightness and area. 
The resulting values range from $(0.3{\rm --}50) \times 10^{47}$\,/\,s.
For comparison, we also extracted the (cumulative) photon rate from a larger region further away from the Galactic plane (see \autoref{fig:HII_region}), and obtained a value of $2.5 \times 10^{48}$\,/\,s.
The resulting expected luminosities range from $10^{34}$ to $10^{35}$\,erg\,/\,s for individual stars. 
For each of the different star groups (O stars only, OB stars only, and both O and OB stars), we obtain luminosities of $\sim 10^{37}$\,/\,s (see \autoref{tab:Qstars} for the results of the different star groupings).

\begin{table}[htbp]
    \centering
     \caption{Expected H$\alpha$ luminosity $L_{\rm exp}$ estimated at the bow-shock region (solid orange box in \autoref{fig:HII_region}) using \autoref{eq:photi1} assuming nearby O and OB stars as photo-ionisation sources. The details of these stars are taken from \citet{2003AJ....125.2531R} and \texttt{Simbad} \citep{2000A&AS..143....9W} and are limited to the region between $15.5^\circ \leq l \leq 21^\circ$ and  $-5^\circ \leq b \leq 5^\circ$ and distances 1 to 3\,kpc.
    Results are also listed for two additional O stars south of the bow-shock, HD\,175754  (O8 II((f)) and  HD\,175876 (O6.5 III(n)(f)) (red circles in \autoref{fig:HII_region}).
     Details of the stars are listed in \autoref{tab:star_photo_fesc} and \autoref{tab:star_photo_fesc2}.
    }
    \begin{tabular}{lcc}
        \toprule
        Star Grouping & \# Stars & ${L_{{\rm exp, }d=2.5 {\rm kpc}}}$ \\
         &  & (erg\,/\,s) \\
        \midrule
        O stars & 33 & $7.2 \times 10^{36}$\\
        OB stars & 103 & $7.2 \times 10^{36}$ \\
        O+OB stars & 136 & $14.4 \times 10^{36}$ \\
        \midrule
        HD 175754 & 1 & $0.4 \times 10^{36}$ \\
        HD 175876 & 1 & $1.3 \times 10^{36}$ \\
        \bottomrule
    \end{tabular}
    \label{tab:Qstars}
\end{table}

Additionally, we calculate the expected luminosity with the two O-type stars HD\,175754 and HD\,175876 as LyC photon sources. 
They are found about 5$^\circ$ south of the Scutum Supershell (see \autoref{fig:HII_region}), and were highlighted by \citet{2000ApJ...532..943C} as potentially influencing the dynamics of the region. 
In this case, $Q_{\rm env}$ is taken again from a circular region of radius $0.2^\circ$.
For both stars, $Q_{\rm env}$ is $\sim 6 \times 10^{46}$\,/\,s using this method, while extracting it from a larger region between the stars and the bow-shock leads to rates of $\sim 2 \times 10^{48}$\,/\,s.
The corresponding luminosities only vary slightly between these two methods and are listed in \autoref{tab:Qstars}.

The expected H$\alpha$ luminosities $L_{\rm exp}$ for the various stars and star groupings are given in \autoref{tab:Qstars}. 
For the O- and OB-star groupings, $L_{\rm exp}$ easily matches the measured H$\alpha$ luminosity of the bow-shock region. 
Similarly, ionisation from the two O stars HD\,175754 and HD\,175876 further south individually could also play a role. 
Overall, photo-ionisation driven by the O and OB\footnote{Due to varied classification criteria, the exact type O or B is not precisely known \citep{2003AJ....125.2531R}.} stars present in OB associations or from individual stars around the field likely provide the bulk of the excitation energy for the H$\alpha$ bow-shock emission. 
For the H$\alpha$ spine feature, which is even closer to these stars by about 100\,pc, the role of photo-ionisation would likely be dominant. 
{Such a conclusion is further supported by the placement of the Galactic worm GW\,16.9$-$3.8, which runs along the spine, with the neutral hydrogen seen to the right of the spine, and the ionising sources on the left. In this scenario, the spine forms the ionised surface of the worm.}

Note that we have neglected reddening in our measured H$\alpha$ luminosities and in determining $Q_{\rm env}$ in the above calculations. 
Using the Galactic Dust Reddening and Extinction online tool\footnote{\url{https://irsa.ipac.caltech.edu/applications/DUST/}} \citep{2011ApJ...737..103S,1998ApJ...500..525S}, we determined an extinction level of A$_{v}=(1.26{\pm0.09})$\,mag in the bow-shock region, increasing to about 2\,mag towards the top end of the spine. 
{However, this converts to an extinction at H$\alpha$, A$_{\textnormal{H}\alpha} = \sim 0.4 $ mag, resulting in an overall increase in luminosity of only 40\%. This also applies to $Q_{\rm env}$..}
Since $Q_{\rm env}$ is somewhat negligible compared to $Q_\star$ for larger groups of stars, and so our overall conclusions remain unchanged.

\subsection{Excitation from Collision and/or Shocks} \label{subsec:E_C_S}

Photo-ionisation appears to be an important excitation mechanism for the H$\alpha$ spine and the bow-shock feature. 
However, given the interesting morphology of the bow-shock and its multi-wavelength counterparts, we will explore the potential role of alternative or additional excitation via shocks or collisions.

A number of studies have noted that \blue{a combination of H$\alpha$, [S\,II], radio continuum, and infrared continuum can be used to help determine the H$\alpha$ excitation processes, 
and important in classifying emissions from SNRs, planetary nebulae, and \HII{} regions (e.g.\, \citet{1992ARA&A..30...11D,1977A&A....60..147S,1993A&AS...98..327B}).}
\blue{Radio continuum emission at levels of 0.001 to 0.7\,Jy/beam, together with H$\alpha$ emission, are used by \citet{stupar2008newly} as an indicator of shock heating from SNRs.
A study by \citet{1989ApJS...70..181A, 1992ApJS...81..715S} points out that certain infrared colour ratios may indicate the origin of excitation, such as from supernova remnants (SNRs), planetary nebulae, and \HII{} regions.}

\blue{
Interestingly, for the right side of the bow-shock, we find infrared colour ratios for 25\,$\mu$m/60\,$\mu$m ($\lesssim 0.1$) and 60\,$\mu$m/100\,$\mu$m ($\sim$ 0.3) that appear to be consistent with those found for old ($>10^4$\,yr) SNRs similar to those of SNR\,G205.5$+$0.5 and others  \citep[see Table 2][]{1992ApJS...81..715S} 
and also with with \HII{} regions, as shown by \citet{1989ApJS...70..181A}. On the other hand, we find no overlapping catalogued \HII{} regions \citep{2014ApJS..212....1A}.
Moreover, we find for this region a peak in [S\,II] emission (\autoref{fig:sii_image}), H$\alpha$ emission (\autoref{V_l}), and overlapping radio continuum (\autoref{Fig:PMN_radio}). Overall, the combination of these multi-wavelength features might suggest some contribution from shock heating.
}

Some further insight into the potential for shock-related influences may come from the recent work of \citet{2022OAst...31..154D}, who conducted a 3D hydrodynamic simulation to study the evolution of supernova (SN) explosions in OB associations \blue{based on a model by \citet{Vasiliev2013}, which did not consider stellar photoionisation and stellar winds}. 
They propose OB associations and their SN explosions as driving mechanisms for large-scale outflow-driven H$\alpha$ emission (see their Figure 9).  
They examined three OB association models: one positioned at the Galactic midplane and the other two positioned at heights of 20\,pc and 60\,pc above the Galactic plane. 
The ages of the OB associations for the three models varied between 10 and 20 Myr. 
SN events within the OB association, which release approximately 10$^{51}$\,erg every 1-2\,Myr, can result in outflows observable in H$\alpha$ and X-rays (0.7--1.2\,keV). 
In one case, for OB associations 20\,pc above the plane and 16\,Myr after the first SN, their predicted H$\alpha$ emission resembles the features we observe and includes quite distinctive central outflow and bow-shock features exceeding surface brightness levels of 100\,R. 
Similar features are predicted for the \HI{} emission. 
For OB associations modelled further (60\,pc) from the plane, features resembling a blow out are predicted for the H$\alpha$ and the \HI{} emissions. 

\serob{} is associated with the massive star cluster NGC\,6611, which includes two populations of stars - a young population with an age of 1.3\,Myr and an old population with an age of 7.5\,Myr (with almost all OB stars belonging to the young population) \citep{Stoop2023}.
It is located $\sim$10\,pc (0.3$^\circ$) above the Galactic plane. 
Ser\,OB2, with an age of 7.8\,Myr \citep{2018MNRAS.473..849D}, is located $\sim40$\,pc (1.5$^\circ$) above the Galactic plane. 
Sct\,OB3 is still young with an age of 1.3\,Myr \citep{2018MNRAS.473..849D}, and is located approximately 15\,pc (0.6$^\circ$) below the Galactic plane. 
All three OB associations have at some point in the past $\sim 1$\,Myr been positionally aligned with an elongation of the H$\alpha$ spine, according to their proper motions (see \autoref{fig:Star-pphotoion}).
These OB associations are key sources of energy that stimulate H$\alpha$ emission, through  SNe, stellar winds, and  photoionisation.

As discussed earlier, SNR\,G18.7--2.2 (potentially linked to \PP{}) and the energetic pulsar PSR\,J1826$-$1334, are both likely at a distance of 4\,kpc or greater. 
The energetic but radio-quiet pulsar PSR\,J1826$-$1256, linked to \PPP{}, has a distance estimate of $\sim 3.5$\,kpc \citep{Karpova:2019}. 
The other SNRs catalogued from \cite{2019JApA...40...36G} all have likely distances of $>4$\,kpc, given their small radio diameters. 
Additionally, we find no other pulsars with distances $<3$\,kpc within the region \citep{manchester2016vizier}.

\citet{2012A&A...543A..26M} suggested that the high-mass X-ray binary LS\,5039, also known as the gamma-ray binary HESS\,J1826$-$148, originated within Sct\,OB3. 
LS\,5039 is at a distance of (2.0$\pm$0.2)\,kpc \citep{2015MNRAS.451...59M,2005Sci...309..746A,2006A&A...460..743A} and hosts a neutron star with a possibly high magnetic field of $\sim 10^{11}$\,T, based on a 9\,s X-ray periodicity observed by \citet{2020PhRvL.125k1103Y} and \citet{2023ApJ...959...79M}. 
This would classify the neutron star as a magnetar, but the 9\,s periodicity has been questioned by \citet{2021ApJ...915...61V} and \citet{2023ApJ...958...79K}. 
Magnetars have been linked to unusually energetic hypernovae (e.g. \citet{1992ApJ...392L...9D}) that may generate asymmetric outflows (such as in long gamma-ray bursts).
\blue{Given the links to hypernovae, and the rarity of magnetars in the Milky Way (only 30 are currently catalogued \citep{2014ApJS..212....6O}),}
LS\,5039 might be an interesting, although speculative, contributor to the dynamics of the Scutum Supershell and associated H$\alpha$ emission. 
The H$\alpha$ luminosities of the spine and bow-shock are in fact several orders of magnitude larger than those seen arising in collisional or shock-driven scenarios, such as protostellar jets, and somewhat similar to those found in the black-hole X-ray binary Cyg\,X-1 and the unique SS-433 micro-quasar system (see \autoref{tab:L_other_source} where we compare these luminosities).
In this scenario, the spine morphology may be influenced by such an outflow. 
However, as discussed earlier, it appears more likely to be guided by the Galactic worm in \HI{} emission (and thus photo-ionisation).

 % stellar wind shocks from individual O or B stars to match the angular scale of H$\alpha$ bow-shock,

\section{Conclusions}

This work has focused on the  H$\alpha$ spectral line emission towards the Scutum Supershell that was first studied by \citet{2000ApJ...532..943C} using the WHAM survey with a $\sim 1^\circ$ angular resolution. 
Towards this region, there are several OB associations, numerous catalogued O-type stars, supernova remnants, pulsar wind nebulae and a high-mass X-ray binary LS\,5039, that could drive the excitation responsible for the H$\alpha$ emission.

Based on recent H$\alpha$ maps from \citet{finkbeiner2003full}, the SuperCOSMOS H$\alpha$ Survey (SHS) \citep{2005MNRAS.362..689P} and our own observations (with a Skywatcher Evolux 62ED refractor telescope), all with arc-second to arc-minute angular resolution, we identify H$\alpha$ emission features, resembling a long spine and a bow-shock that each extend over several degrees in length. 
Our spectral analysis of the H$\alpha$ data from WHAM suggests that these features have a kinematic distance in the range from 2 to 3\,kpc.

These multi-wavelength observations, alongside the presence of OB associations, indicate significant energetic processes in the region.
Our observations revealed that areas within the bow-shock with the brightest H$\alpha$ emission also displayed [S\,II] emission. 
Additionally, the same region exhibits continuum radio emission at 4.85\,GHz, as well as infrared emission at 60$\mu$m and 100$\mu$m. 
These findings could indicate the influence of shock heated gas in shaping the characteristics of the bow-shock.

We also examined \HI{} data from the SGPS and found broadly similar blowout features consistent with those found by \citet{2000ApJ...532..943C} in earlier \HI{} data. 
Moreover, a comparison of the H$\alpha$ data with the SGPS \HI{} data cube, integrated over the 20 to 30\,km\,/\,s range (2--3\,kpc distance), reveals a clear void in the H$\alpha$ emission overlapping the Galactic worm GW\,16.9$-$3.8 \HI{} emission running along a North-South direction adjacent to the spine. 
Thus, the Galactic worm appears to be playing a physical role in shaping the morphology of the H$\alpha$ spine.

Our analysis of recent MAXI SSC X-ray emission in the soft (0.7--1.0\,keV) band revealed a similar picture as shown by \citet{2000ApJ...532..943C} using ROSAT\,PSPC data in the (0.4--1.2\,keV) range. 
The X-ray emission is found inside the boundary of the Scutum Supershell and partly overlaps the H$\alpha$ features. 
The X-ray emission in a wider field ($40^\circ \times 40^\circ$) reveals several X-ray peaks further to the Galactic-south, which may be connected with larger--scale Galactic diffuse X-ray emission. 
Overall, the X-ray emission is consistent with hot gas over large regions of the Scutum Supershell. 

We considered both photo-ionisation and shock-heating as sources of excitation for the H$\alpha$ emission. 
Following the method of \citet{1997ApJ...474L..31D}, we estimated the H$\alpha$ luminosity via Lyman continuum photons from O- and B-type stars associated with the three nearby OB associations (\serob{}, Ser\,OB2 and Sct\,OB3), and additional O- and B-type stars from a wider region. 
Assuming a distance of 2.5\,kpc for the H$\alpha$ features, the OB associations and/or the wider grouping of O and B stars could provide sufficient excitation for photo-ionisation to match the measured H$\alpha$ luminosity (few $10^{36}$\,erg\,/\,s). 
If the H$\alpha$ spine and bow-shock features are located at the same 1.5\,kpc distance as determined for the OB association distances (1.3 to 1.7\,kpc; based on GAIA data), the predicted photo-ionisation luminosity from the OB associations increases by a factor 3 to 6, further boosting this interpretation.

For the bow-shock H$\alpha$ feature however, its overlap with optical [S\,II] and continuum emission in the radio and infrared bands could suggest some additional shock-induced excitation or heating. 
Due to the lack of an absolute flux calibration for our [S\,II] image, we cannot presently determine the [S\,II]/H$\alpha$ ratio,  an important diagnostic of the potential for shock-induced excitation. 
Future work will include an accurate determination of the [SII]/H$\alpha$ ratio using WiFes \citep{2007Ap&SS.310..255D} spectroscopic observations towards the bow-shock and spine regions.

Recent MHD simulations by \citet{2022OAst...31..154D} demonstrate the potential for supernova events to drive outflow and bow-shock types of features of the same energetic nature and physical scale as the H$\alpha$ emission we observe here.
The nearby $\gamma$-ray sources \PP\, and \PPP\, are at distances $>$4\,kpc and so are likely to be unrelated.
However, the high-mass X-ray binary LS\,5039 (at 2\,kpc distance, and linked to Sct\,OB3) has been suggested to host a magnetar that could have been produced in a hypernova event with outflows (akin to a long gamma-ray burst). 
We could speculate that such a hypernova outflows might have influenced the dynamics of the H$\alpha$ emission, although it is clear, from our calculations, that photo-ionisation plays an important role. 

We could speculate that such a hypernova outflows might have influenced the dynamics of the H$\alpha$ emission, although it is clear from our calculations that photoionization plays an important role.

\section*{Acknowledgements}
We acknowledge the Southern H-Alpha Sky Survey Atlas (SHASSA), which is supported by the National Science Foundation. The Wisconsin H$\alpha$ Mapper and its H$\alpha$ Sky Survey have been funded primarily by the National Science Foundation. The facility was designed and built with the help of the University of Wisconsin Graduate School, Physical Sciences Lab, and Space Astronomy Lab. NOAO staff at Kitt Peak and Cerro Tololo provided on-site support for its remote operation. This research has made use of MAXI data provided by RIKEN, JAXA, and the MAXI team.  This research made use of Astropy,\footnote{http://www.astropy.org} a community-developed core Python package for Astronomy \citep{astropy:2013, astropy:2018}.
We sincerely thank the reviewers for their time, effort, and valuable feedback, which greatly improved our manuscript.
\bibliography{bib/bibll.bib}

%%%%%%%%%%%%%%%%%%%%%%%%%%%%%%%%%%%%%%%%%%%%%%%%%%
%%%%%%%%%%%%%%%%% APPENDICES %%%%%%%%%%%%%%%%%%%%%

\counterwithin{figure}{section}
\counterwithin{equation}{section}
\counterwithin{table}{section}

\appendix
\clearpage

\section{Data Analysis}

\subsection{H$\alpha$ Emission}

\begin{figure}[!h]
\centering
\includegraphics[width=\textwidth]{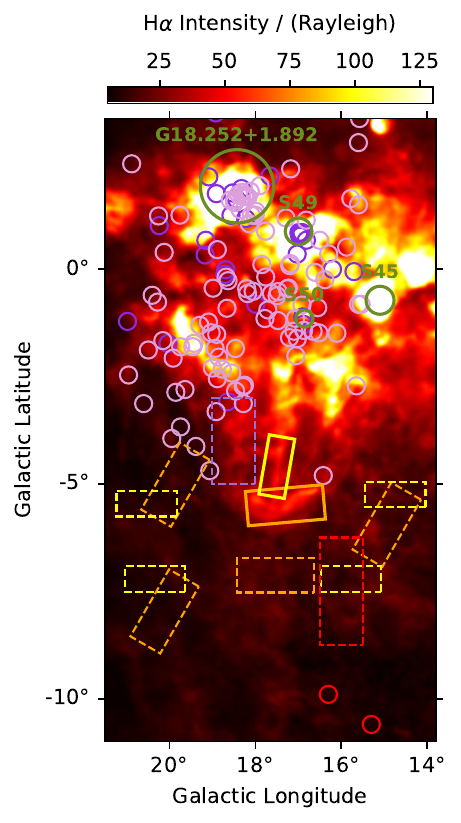}
	\caption{Illustration of several regions used throughout our analyses to extract H$\alpha$ intensities from \citet{finkbeiner2003full} and WHAM. The rectangular regions (corresponding to \autoref{tab:Bakcground_ha}) include the spine (solid yellow) and the bow-shock (solid orange) features and their corresponding backgrounds (dashed). The purple regions (circles and rectangle) were used to extract the photon rates $Q_{\rm env}$ in \autoref{eq:photi1} for O stars (dark purple) and OB stars (light purple), and the red regions (circles and rectangle) for the two HD stars. The \HII{} regions are shown as well. } 
	\label{fig:HII_region}
\end{figure}

To calculate the H$\alpha$ luminosities $L_{{\rm H} \alpha}$ of the bow-shock and spine features reported in \autoref{sec:luminosity}, we extracted intensities from regions surrounding them and from regions to estimate the corresponding background to be subtracted. These regions are defined in \autoref{tab:Bakcground_ha} and illustrated in  \autoref{fig:HII_region}. The table also provides the resulting luminosities, extracted from both Finkbeiner and WHAM maps. 
Comparisons to luminosities of other sources are listed in \autoref{tab:L_other_source}.

\begin{table}
    \centering
    \caption{Definition of rectangular regions and resulting H$\alpha$ luminosities ($L_{{\rm H}\alpha}$) for Finkbeiner and WHAM surveys. These results were used in \autoref{sec:luminosity}. Signal regions are highlighted in bold, while the remaining ones are background regions. The rectangle is defined with width $w$ in North-South and height $h$ in East-West direction if the angle $\theta$ is 0. Galactic longitude and latitude are $l$ and $b$.}
    \begin{tabular}{cccccccc}
        \toprule
         $ {l}$ &  $ {b}$ & $w$ & $h$ & $\theta$ & $L_{{\rm H}\alpha{\textnormal{\tiny , Finkb.}}}$ & $L_{{\rm H}\alpha{\textnormal{\tiny , WHAM}}}$\\
        \scriptsize(deg) & \scriptsize(deg) & \scriptsize(deg) & \scriptsize(deg) & \scriptsize(deg) & \scriptsize ($10^{36}$ erg/s) & \scriptsize ($10^{36}$ erg/s) \\
        
        \midrule
        \multicolumn{7}{c}{Bow shock }\\
        \bfseries 17.30 & \bfseries -5.50 & \bfseries 0.8 & \bfseries 1.8 & \bfseries 5 & \bfseries 3.18 & \bfseries 2.42 \\
        17.53 & -7.13 & 0.8 & 1.8 & 0 & 1.31 & 1.30 \\
        14.94 & -5.95 & 0.8 & 1.8 & 60 & 1.39 & 1.29 \\
        19.86 & -5.02 & 0.8 & 1.8 & 60 & 0.92 & 1.01 \\
        20.11 & -7.97 & 0.8 & 1.8 & 60 & 0.84 & 0.85 \\
        
        \midrule
        \multicolumn{7}{c}{Spine}\\
        \bfseries 17.50 & \bfseries -4.60 & \bfseries 0.6 & \bfseries 1.4 & \bfseries 80 & \bfseries 1.86 & \bfseries 1.41 \\
        14.74 & -5.25 & 0.6 & 1.4 & 0 & 0.80 & 0.81 \\
        15.78 & -7.22 & 0.6 & 1.4 & 0 & 0.72 & 0.74 \\
        20.53 & -5.46 & 0.6 & 1.4 & 0 & 0.48 & 0.50 \\
        20.33 & -7.22 & 0.6 & 1.4 & 0 & 0.50 & 0.48 \\
        \bottomrule
    \end{tabular}
    \label{tab:Bakcground_ha}
\end{table}

We compared our obtained H$\alpha$ luminosities \autoref{tab:Bakcground_ha} to the ones of other sources \autoref{tab:L_other_source}.

\begin{table}
    \centering
    \caption{H$\alpha$ luminosities ($L_{{\rm H}\alpha}$) of various sources as comparison to the ones of the spine and bow--shock features.}
    \begin{tabular}{lcc}
        \toprule
        Object & Luminosity & Ref. \\
            &   (erg/s) & \\
        \midrule
        M82 halo & 10$^{40}$ &                                 \citet{1998ApJ...493..129S}      \\%M87 has ~ 9000 light year 2.7 kpc jet length
        SS\,433 & $< 10^{39}$ &                       \citet{10.1007/978-94-011-1924-5_34}       \\ % distance 180 pc  in the text the ref
        Cygnus X--1 & ${6}\times$10$^{34}$&           \citet{2007MNRAS.376.1341R}      \\%distance 3 pc  (Jet–ISM Interactions near the Microquasars GRS1758−258 and 1E 1740.7−2942)
        HH34 & 10$^{31}$ &                                      \citet{1988AA...200...99B}   \\%distance 0.24 pc
        \bottomrule
    \end{tabular}
\label{tab:L_other_source}
\end{table}

In \autoref{sec:photoion} we performed a dedicated analyses of photo-ionisation from OB associations in the region.
\autoref{tab:ob_assoc} provides a list of all OB associations \citep{2020NewAR..9001549W} towards the Scutum Supershell, together with their associated star clusters (references in table).

\begin{table*}
    \centering
    \caption{OB associations Galactic longitude and latitude are $l$ and $b$, respectively. Diameters are represented $dl$ and $db$ are from \citet{1995Ast}. Star clusters associated with OB associations are taken from \citet{2020NewAR..9001549W}. Individual star cluster data include age, proper motions in right ascension PM$_{\rm ra}$, and declination PM$_{\rm dec}$, along with their references.}
    \begin{tabular}{lcccc|ccccc}
        \toprule
        Name & $l$ & $b$ & $dl$ & $db$ & Star Cluster & Age & PM$_{\rm ra}$ & PM$_{\rm dec}$ & Ref. \\
        & (deg) & (deg) & (deg) & (deg) & & (Myr) & (mas\,/\,yr) & (mas\,/\,yr) & \\
        \midrule

        Ser\,OB1B & 16.93 & 0.76 & 0.2 & 0.3 & NGC\,6611 & 1.3 (7.5) & 0.21 & -1.59 & \citet{Stoop2023} \\
        Ser\,OB2 & 18.23 & 1.66 & 0.9 & 1.3 & NGC\,6604 & 7.8 & 1.54 & -7.31 & \citet{2017AstBu..72..257L} \\

        Sct\,OB3 & 17.23 & -0.84 & 1.0 & 1.3 & Sct\,OB3 & 1.3 & -0.98 & -1.84 &\citet{Kharchenko2013} \\
        \bottomrule
    \end{tabular}
\label{tab:ob_assoc}
\end{table*}

\subsection{H$\alpha$ Spectra} \label{Gpy}

\begin{figure}
    \centering 
    \includegraphics[width=0.95\textwidth]{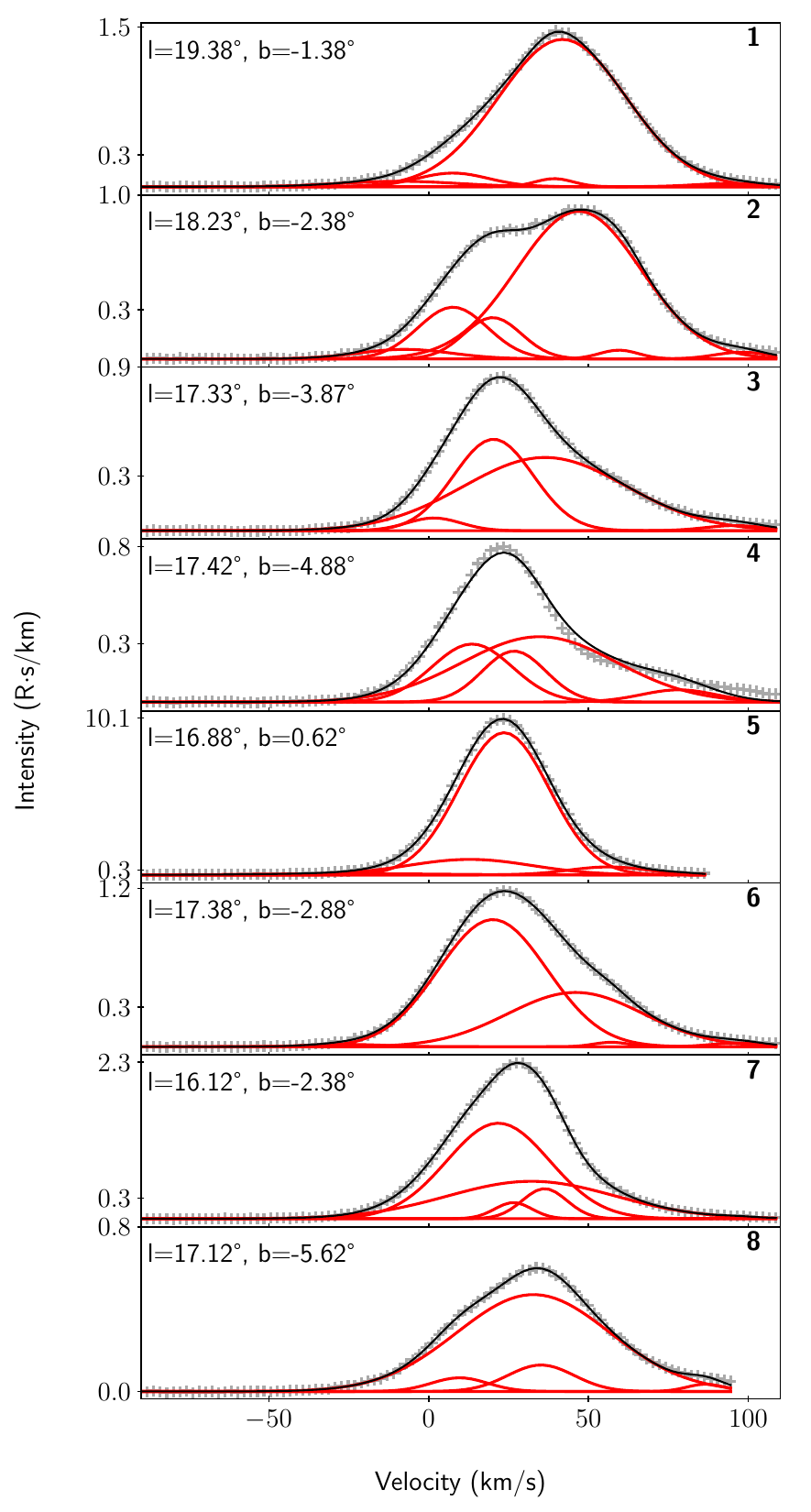} 
    \caption{WHAM H$\alpha$ spectra from  \citet{1998PASA...15...14R} for the eight pixels shown in \autoref{V_l}. The individual Gaussian components (derived with {\tt GaussPy+}) are shown in red and the total in black.}
    \label{fig:gausspy}
\end{figure}

We used the {\tt GaussPy+} package \citep{2019A&A...628A..78R,2015AJ....149..138L} to decompose the components of the WHAM H$\alpha$ spectra for the region shown in \autoref{V_l}.
The resulting Gaussian components for the eight spectra shown in  \autoref{V_l} are shown in \autoref{fig:gausspy} .
In our case, the {\tt GaussPy+} procedure decomposed each pixel's spectrum into typically five to six Gaussian components with one to two clearly dominant ones. 
In the {\tt GaussPy+} process, we trained the smoothing parameter $\alpha$ using a wide region with boundaries $b=-8^\circ$\,to\,8$^\circ$ and $l=10^\circ$\,to\,24$^\circ$.
We trained with 250 spectra, resulting in an $\alpha$ value of 4.68.
Due to the spectral  resolution of WHAM, we have set a minimum FWHM of 12\,km\,s$^{-1}$ (6 channels) for each Gaussian component. 

\subsection{Infrared Emission}

\begin{figure}[!h]
    \centering
    \includegraphics[width=\textwidth]{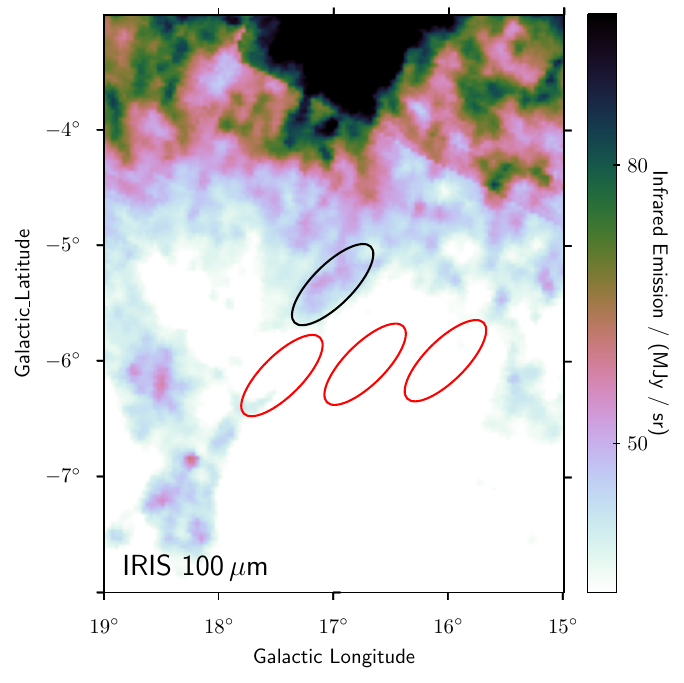}
    \caption{IRIS 100 $\mu$m observations overlaid with regions used to extract the emission for the bow-shock (black ellipse) and backgrounds (red  ellipses).}
    \label{fig:IRASbk}
\end{figure}

The IRIS infrared (IR) emission from \citet{2005ApJS..157..302M} shows significant emission in the 60$\mu$m and 100$\mu$m images towards the right bow-shock region.
We estimated the IR flux in these two wavelengths in \autoref{sec:IRIS}.
The corresponding region together with the regions used as background estimates are listed in \autoref{tab:IRIS_bk} (and illustrated in \autoref{fig:IRASbk}), together with the obtained fluxes.

\begin{table}[!ht]
    \centering
    \caption{IRIS infrared fluxes $F_{\mu {\rm m}}$ for different regions (bold refers to the bow-shock region). The ellipse is defined with diameter $w$ of horizontal axis, diameter $h$ of vertical axis, and rotation angle $\rho$ (anti-clockwise). Galactic longitude and latitude are $l$ and $b$.}
    \label{tab:IRIS_bk}
    \begin{tabular}{cccccccc}
    \toprule
     ${l}$  & ${b}$ & $w$ & $h$ & $\rho$ & $F_{100}$ & $F_{60}$  & $F_{25}$ \\
     (deg)    & (deg)       &      (deg)  &  (deg)    &  (deg)   &    (Jy)                &    (Jy)             &  (Jy)             \\
    \midrule
    \bfseries 17.00 & \bfseries -5.34 & \bfseries 0.35 & \bfseries 0.93 & \bfseries 45 & \bfseries 171 & \bfseries 56.7 & \bfseries 8.1 \\
    16.02& -5.99 & 0.35 & 0.93 &   45 & 115 & 33.6 & 7.2 \\
    16.72& -6.02 & 0.35 & 0.93 &   45 & 108 & 31.3 & 7.0 \\
    17.44& -6.12 & 0.35 & 0.93 &   45 & 107 & 32.8 & 7.2\\
    \bottomrule
\end{tabular}
\end{table}

\subsection{X-ray Emission}

\begin{table*}
    \centering
    \caption{Absorbed and unabsorbed fluxes ($F_{\rm  X-ray,  A}$, $F_{\rm  X-ray}$) and luminosities ($L_{\rm  X-ray,  A}$, $L_{\rm  X-ray}$) for different temperatures $T$. The values were obtained with PIMMS for the region marked (dashed rectangle) in \autoref{fig:x-r} for both ROSAT\,PSPC (0.4-1.2\,keV) and MAXI\,SSC (0.7-1.0\,keV) X-ray observations, assuming a distance of 2.5\,kpc.
    An average column density of N$_{\rm H}$ = 5.8\,$\times$\,10$^{21}$\,/\,cm$^{2}$ was assumed. 
    } 
    \begin{threeparttable}
    \begin{tabular}{l|cccc|cccc}
        \toprule
        \  & \multicolumn{4}{c|}{  MAXI\,SSC  (0.7-1.0 keV)  }   &  \multicolumn{4}{c}{ROSAT\,PSPC (0.4-1.2 keV) }   \\
       $T$ &   $F_{\rm  X-ray,  A}$ & $L_{\rm  X-ray,  A}$ &  $F_{\rm  X-ray}$  & $L_{\rm  X-ray}$  &  $F_{\rm  X-ray,  A}$  & $L_{\rm  X-ray,  A}$  &  $F_{\rm  X-ray}$  & $L_{\rm  X-ray}$ \\
       (K) &  (erg\,/\,s \,/\, cm$^{2}$)    & (erg\,/\,s)   &   (erg\,/\,s \,/\, cm$^{2}$)    & (erg\,/\,s)    &   (erg\,/\,s \,/\, cm$^{2}$)    & (erg\,/\,s)   &   (erg\,/\,s \,/\, cm$^{2}$)    & (erg\,/\,s)   \\ 
       \midrule
        $10^7$ &         1.5$ \times 10^{-9}$ & 1.1$ \times 10^{36}$ &           6.1$ \times 10^{-9}$ &             4.6$ \times 10^{36}$ &           1.5$ \times 10^{-9}$ &             1.1$ \times 10^{36}$ &             5.3$ \times 10^{-9}$ &               3.9$ \times 10^{36}$ \\
         $10^{6.5}$ &         1.5$ \times 10^{-9}$ & 1.1$ \times 10^{36}$ &           6.6$ \times 10^{-9}$ &             5.0$ \times 10^{36}$ &           1.5$ \times 10^{-9}$ &             1.1$ \times 10^{36}$ &             7.1$ \times 10^{-9}$ &               5.3$ \times 10^{36}$ \\
           $10^6$ &         1.8$ \times 10^{-9}$ & 1.3$ \times 10^{36}$ &           9.5$ \times 10^{-9}$ &             7.1$ \times 10^{36}$ &           2.0$ \times 10^{-9}$ &             1.5$ \times 10^{36}$ &             4.2$ \times 10^{-8}$ &               3.1$ \times 10^{37}$ \\
        \bottomrule
    \end{tabular}
    \end{threeparttable}
    \label{tab:X-ray_pimf} 
\end{table*}

To calculate the X-ray luminosity similarly to \citep{2000ApJ...532..943C} shown in \autoref{tab:X-ray_pimf}, we used the region at Galactic longitude of 17.1$^\circ$ and latitude of -3.5$^\circ$ to find the X-ray counts and average total column density. 
We subtracted the background counts from the region of interest using the rectangular background regions listed in \autoref{tab:maxi_bkgd}.

\begin{table}
  \caption{Definition of rectangular regions used for the analyses of MAXI\,SSC and ROSAT\,PSPC observations. These results were used in \autoref{sec:luminosity}. The signal region is highlighted in bold, while the remaining ones are background regions. The rectangle is defined with width $w$ in North-South and height $h$ in East-West direction if the angle $\theta$ is 0. Galactic longitude and latitude are $l$ and $b$.}
    \centering  
    \begin{tabular}{ccccc}
    \toprule
    ${l}$ &  $ {b}$   & $w$ & $h$ & $\theta$ \\
    (deg) & (deg) & (deg) & (deg) & (deg) \\
    \midrule
    \bfseries 17.1 & \bfseries  -3.5 & \bfseries  3.7 & \bfseries  4.5 &  \bfseries 2.0 \\
    26.3 &-10.6 & 3.7 & 4.5 & 0.1 \\
    21.9 &-9.9 & 3.7 & 4.5 & 0.1 \\
    \bottomrule
    \end{tabular}
    \label{tab:maxi_bkgd}
\end{table}

\section{Observations}

\subsection{X-ray}

\begin{figure*}[!h]
	\centering 
    \includegraphics[width=0.99\textwidth]{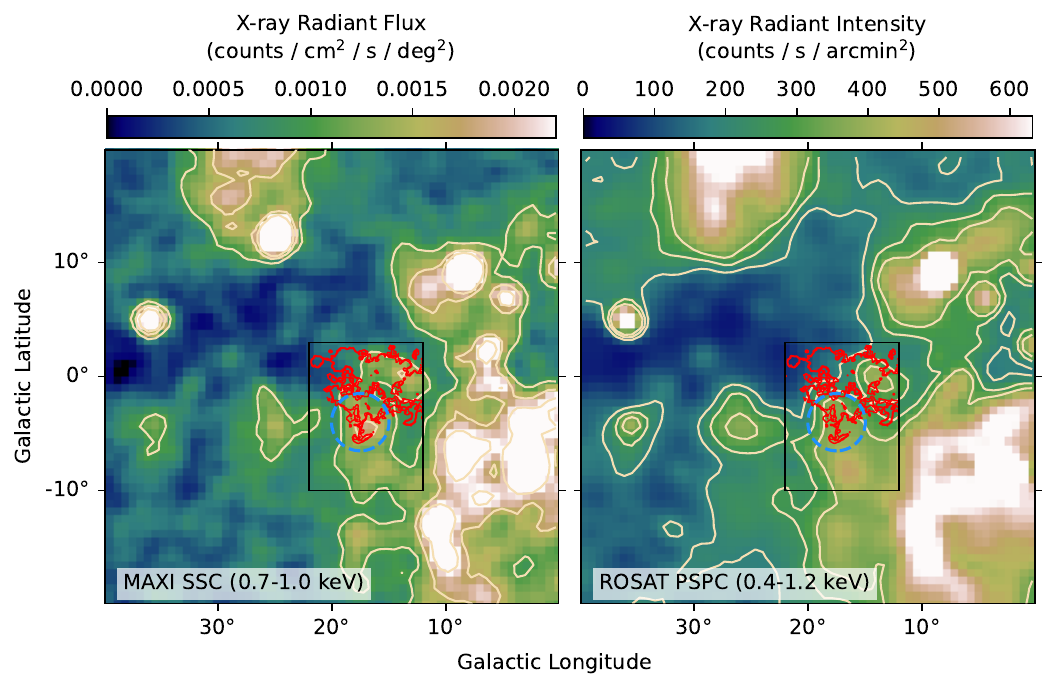}
	\caption{Soft X-ray maps in a wider region around the Scutum supershell (dashed blue circle). Emission from MAXI\,SSC (left; \citet{2014SPIE.9144E..1OM}) for energies 0.7--1.0\,keV is compared with ROSAT\,PSPC (right; \citet{1997ApJ...485..125S}) for energies 0.4--1.2\,keV. Brown contours represent X-ray emission at 5, 10, and 15$\sigma$ level, and red contours H$\alpha$ emission at 40 and 50 Rayleigh \citep{finkbeiner2003full}. The black box indicates the region shown in the other X-ray images (\autoref{fig:x-soft}, \autoref{fig:x-hard}).} 
	\label{fig:xray-wide}
\end{figure*}

\autoref{fig:xray-wide} present wide-field views of the MAXI\,SSC and ROSAT\,PSPC X-ray emission. The MAXI\,SSC image is smoothed with a Gaussian with a standard deviation of 1 pixel (10\,arcmins in length). 

\begin{figure*}[!h]
    \centering 
	\includegraphics[width=0.9\textwidth]{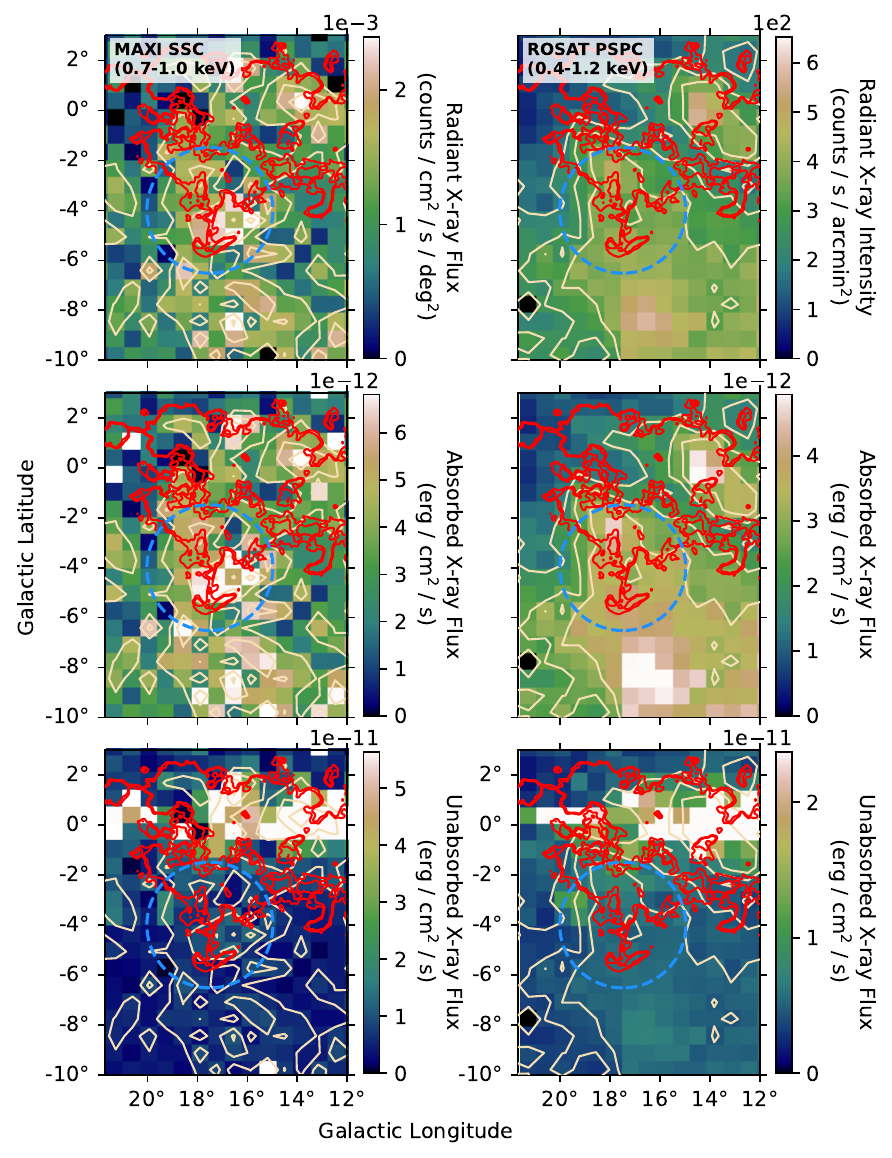}
	\caption{Soft X-ray emission from MAXI~SSC (left; \citet{2014SPIE.9144E..1OM}) and ROSAT\,PSPC (right; \citet{1997ApJ...485..125S}) towards the Scutum supershell (blue circle). We compare X-ray intensities (top), absorbed fluxes (middle) and unabsorbed fluxes (bottom), and used PIMMS to correct for photoelectric absorption. X-ray contours (brown) are shown at 5, 10, and 15\,$\sigma$ for both MAXI\,SSC (0.7-1.0\,keV) and ROSAT\,PSPC (0.4-1.2\,keV). The temperature assumed in the PIMMS model for both analyses is $T=10^7$\,K. The H$\alpha$ contours (red) are shown at 40 and 50\,Rayleigh.}
	\label{fig:x-soft}
\end{figure*}

\begin{figure*}[!h]
    \centering 
	\includegraphics[width=0.9\textwidth]{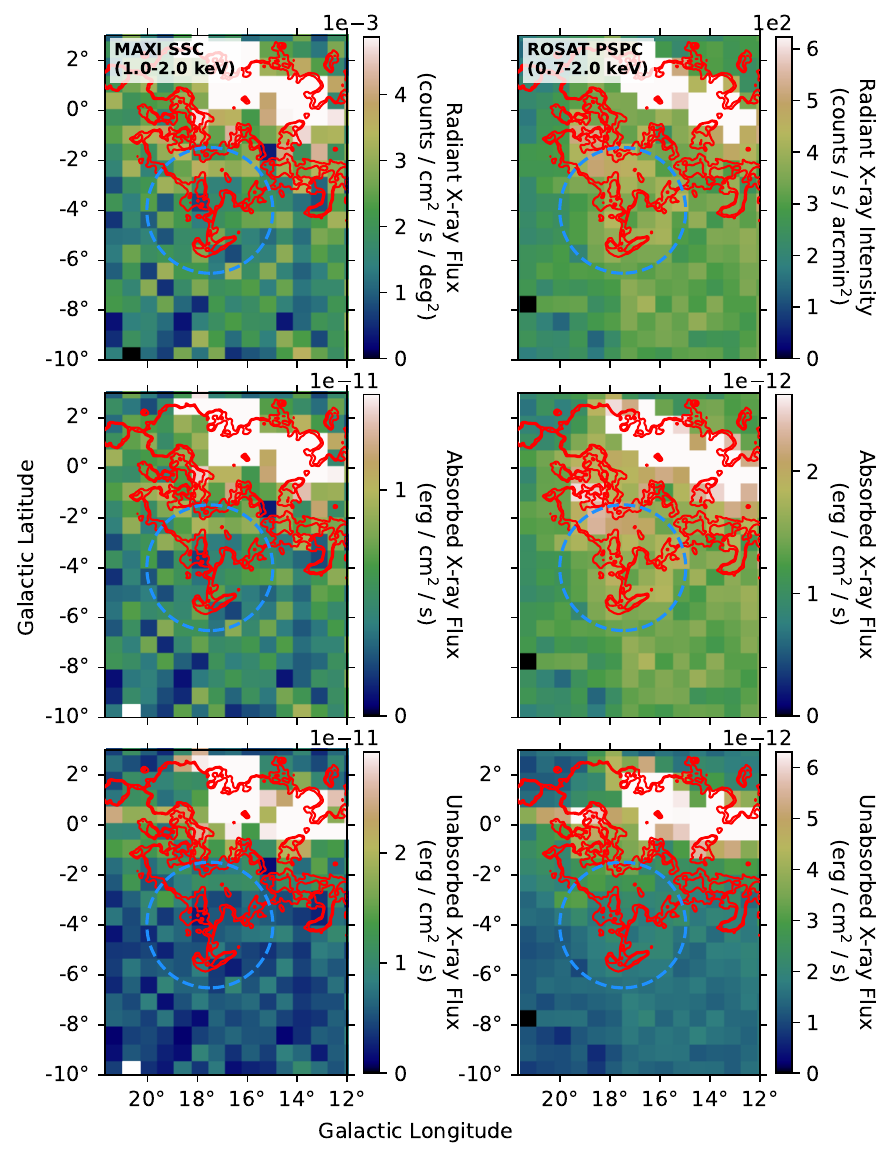}
	\caption{Hard X-ray emission from MAXI~SSC (left; \citet{2014SPIE.9144E..1OM}) and ROSAT\,PSPC (right; \citet{1997ApJ...485..125S}) towards the Scutum supershell (blue circle). We compare X-ray intensities (top), absorbed fluxes (middle) and unabsorbed fluxes (bottom), and used PIMMS to correct for photoelectric absorption. X-ray contours (brown) are shown at 5, 10, and 15\,$\sigma$ for both MAXI\,SSC (1.0-2.0\,keV) and ROSAT\,PSPC (0.7-2.0\,keV). The temperature assumed in the PIMMS model for both analyses is $T=10^7$\,K. The H$\alpha$ contours (red) are shown at 40 and 50\,Rayleigh.}
	\label{fig:x-hard}
\end{figure*}

In \autoref{fig:x-soft} and \autoref{fig:x-hard}, we compare MAXI~SSC (left) and ROSAT\,PSPC (right) X-ray intensities (top), absorbed fluxes (middle) and unabsorbed fluxes (bottom). We used PIMMS to correct for photoelectric absorption as discussed in \autoref{sec:x-ray}.
The high flux regions in the figure are associated with the \HII{} regions G018.426+1.922 and G016.993+0.873 from \citet{2014ApJS..212....1A}.

\subsection{Molecular Hydrogen}

\begin{figure*}
	\centering 
    \includegraphics[width=1\textwidth]{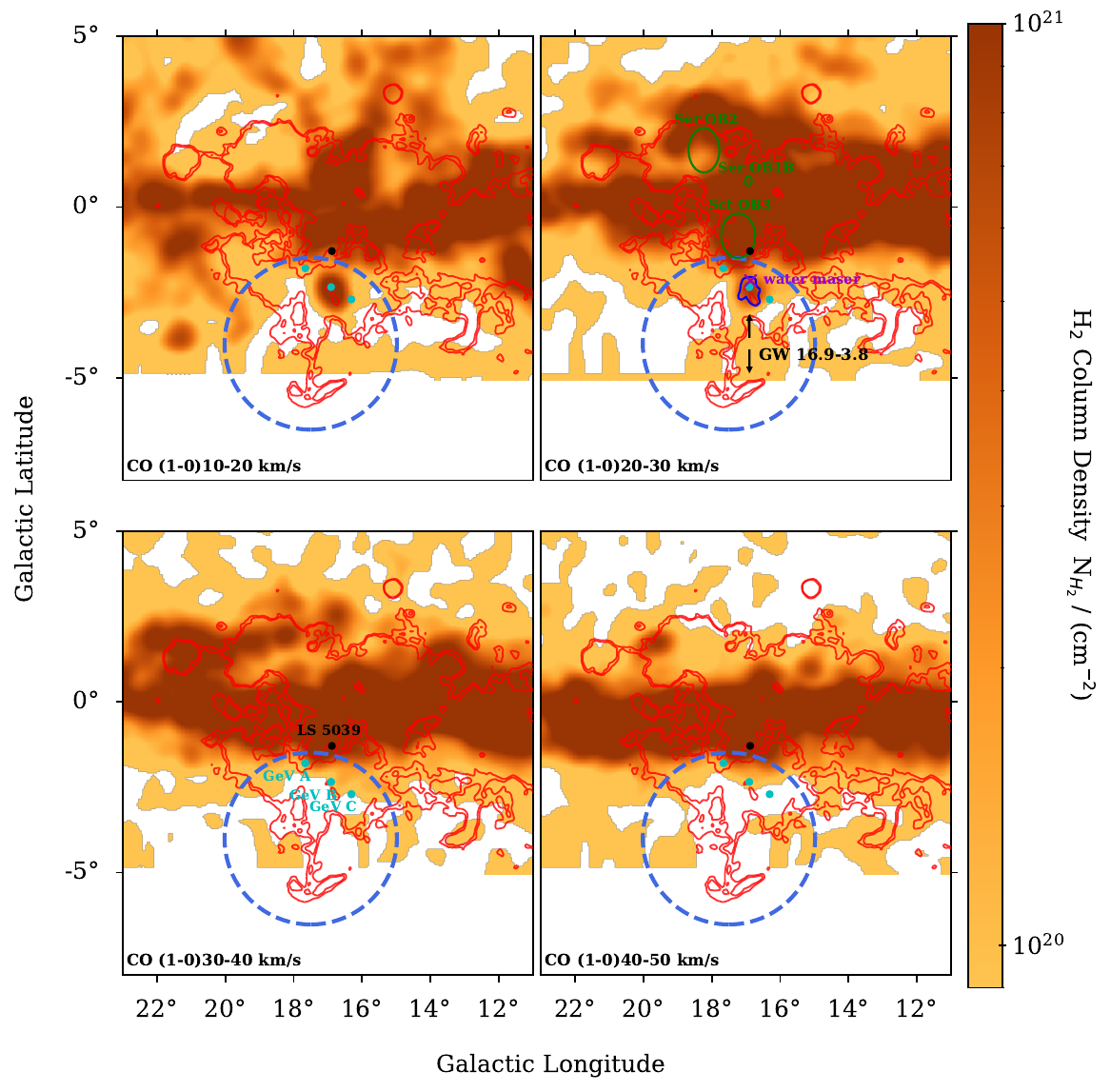}
	\caption{ H$_{2}$ column density maps using $^{12}$CO ($J=$1--0) emission from \citet{2001ApJ...547..792D}, integrated over different velocity ranges. H$\alpha$ emission at 40 and 50\, Rayleigh (red) \citep{finkbeiner2003full} is indicated as black contours, while the	Scutum Supershell boundary is shown as a  blue dashed circle \citep{2000ApJ...532..943C}. 
	GeV\,A, GeV\,B, and GeV\,C (cyan dots) are {\em Fermi}-LAT GeV gamma-ray sources \citep{2019MNRAS.485.1001A}, and LS\,5039 (black dot) is a high-mass X-ray binary featuring high-energy gamma-ray emission.
	The 20--30\,km\,s$^{-1}$ map includes the Nanten\,CO cloud (dark blue contour) for the 15--30\,km\,s$^{-1}$ range from \citet{2021MNRAS.504.1840C}, and the water maser G016.8689$-$02.1552 (purple cross) \citep{2011MNRAS.418.1689U}. The Galactic worm GW\,16.9$-$3.8 is indicated by black arrows \citep{1992ApJ...390..108K}.}
	\label{fig:COmap}
\end{figure*}

We used \citet{2001ApJ...547..792D} $^{12}$CO observations to understand the molecular clouds toward the Scutum Supershell.
\autoref{fig:COmap} shows the $^{12}$CO(1--0) map (converted to H$_2$ column density), covering different velocity ranges.
The CO maps only cover $|b|<5^\circ$, so that the H$\alpha$ bow-shock region is not encompassed. 
The Nanten CO emission (15-30\,km/s) from \citet{2004ASPC..317...59M} encompasses the water maser G016.8689$-$02.1552   \citep{2011MNRAS.418.1689U}, and the gamma-ray source GeV-B towards the north of the Galactic worm GW\,16.9--3.8 and south of the high-mass X-ray binary LS\,5039 \citep{1992ApJ...390..108K}.

\section{Catalogues}

\subsection{Atomic Hydrogen Supershells}

\autoref{tab:obGSH} lists the Galactic supershells so far identified with total energies above $10^{52}$\,erg to compare the Scutum supershell with other supershells. 

\begin{table}[!ht]
    \caption{List of Galactic supershells (GSH) with total energy ($E$) above $10^{52}$\,erg, age, mass, and distance ($D$). Data are from \cite{2019yCat..36240043S}, \cite{2001A&A...374..682E}~($\ast$), \cite{2006ApJ...638..196M}~($\star$), \cite{2011ApJ...741...85D}~($\blacklozenge$), and \cite{2014A&A...562A..69K}~($\blacktriangle$). The Scutum Supershell is highlighted in boldface.}
    \centering
    \small
    \begin{tabular}{lcccc}
    \toprule
    GSH & log($E$) & Age & Mass & $D$ \\
    & (erg) & (Myr) & ($10^{4}$ M$_\odot$) & (kpc) \\ 
    \midrule
    016-01+71 & 52.4 & ~ & 5.8 & 6.3 \\ 
    \textbf{018-04+44} & \textbf{52.0} &  & \textbf{62} & \textbf{3.3}\\ 
    022+01+139 & 53.0 & ~ & 6.4 & 9.5 \\ 
    029+00+133 & 52.6 & ~ & 6.3 & 8.7 \\ 
    033+06-49 & 52.9 & ~ & 6.8 & 22 \\ 
    034-06+65 & 52.1 & ~ & 6.0 & 4.6 \\ 
    041+01+27 & 52.9 & ~ & 6.7 & 2.0 \\ 
    046-01+83 & 52.0 & ~ & 6.0 & 6.9 \\ 
    052+07+39 & 52.1 & ~ & 6.0 & 3.1 \\ 
    057+01-33 & 53.6 & ~ & 7.0 & 13.8 \\ 
    061+00+51 & 52.1 & ~ & 5.7 & 4.8 \\ % The H I supershell GS061+00+51 and its neighbours 
    064-01-97 & 53.8 & ~ & 7.1 & 16.9 \\
    071+06-135 & 54.2 & ~ & 7.8 & 21.6 \\ 
    075-01+39 & 52.9 & ~ & 6.2 & 2.6 \\ 
    081-05-37 & 52.9 & ~ & 6.9 & 7.5 \\ 
    088+02-103 & 54.1 & ~ & 7.3 & 12.6 \\ 
    091-04-69 & 52.8 & ~ & 6.8 & 9 \\ 
    091+02-101 & 52.3 & ~ & 6.3 & 12.3 \\ 
    095+04-113 & 53.5 & ~ & 7.3 & 12.9 \\ 
    103+05-137 & 53.4 & ~ & 7.0 & 15.6 \\ 
    108-04-23 & 52.7 & ~ & 6.1 & 2.5 \\ 
    117-07-67 & 52.2 & ~ & 6.0 & 6.3 \\ 
    123+07-127 & 53.3 & ~ & 7.4 & 15.1 \\ 
    128+01-105 & 52.2 & ~ & 6.3 & 11.4 \\ 
    139-03-69 & 54.8 & ~ & 8.2 & 7.1 \\ 
    20+05+23 & 52.2 & ~ & 6.0 & 4.8 \\ 
    223-02+35 & 52.6 & ~ & 6.6 & 3.5 \\ 
    224+03+75 & 53.4 & ~ & 7.0 & 7.6 \\ 
    242-03+37 & 54.2 & ~ & 7.5 & 3.6 \\ 
    242-03+37$^\star$ & 53.0 & 21 & 4.8 & 1.6 \\ %EVIDENCE FOR CHIMNEY BREAKOUT IN THE GALACTIC SUPERSHELL GSH 24203+37
    277+00+36$^\blacklozenge$ & 53.0 & >10 &  >100 & 6.5 \\ %Molecular Clouds in Supershells: A Case Study of Three Objects in the Walls of GSH 287+04-17 and GSH 277+00+36
    287+04-17$^\blacklozenge$ & 52.0 & >10 &  >100 & 2.6 \\ %Molecular Clouds in Supershells: A Case Study of Three Objects in the Walls of GSH 287+04-17 and GSH 277+00+36
  \bottomrule
    \end{tabular}
    \label{tab:obGSH}
\end{table}

\subsection{O and OB Stars}
OB Stars from \citet{2003AJ....125.2531R} were used, employing {\tt Astropy} and {\tt SIMBAD} to determine the spectral type and distance of each star.
\autoref{tab:star_photo_fesc} lists all O-type stars in the region of interest.
The distance of each star was determined using Gaia DR3 parallax \cite{2020yCat.1350....0G}.
We calculate the distance $r$ to the bow-shock, using the coordinates Galactic longitude of 17.3$^{\circ}$ and Galactic latitude of $-$5.5$^{\circ}$ (\autoref{fig:Star-pphotoion}).
Similarly, \autoref{tab:star_photo_fesc2} lists stars with spectral type `OB' (uncertain type of O or B).

\begin{onecolumn}
\begin{longtable}{lcccccccc}
\caption{List of O-type stars from \citet{2003AJ....125.2531R}. Their spectral type was found using {\tt Astropy} and {\tt SIMBAD}, and the stars are limited to the region between $15.5^\circ \leq l \leq 21^\circ$ and  $-5^\circ \leq b \leq 5^\circ$. The stars' magnitudes (mag) are from \cite{2003AJ....125.2531R} (mag). The distances to Earth (based on Gaia DR3 parallax measurements; \citet{2020yCat.1350....0G}) and the distances $r$ to the bow-shock (GL=17.3$^{\circ}$ and GB=$-$5.5$^{\circ}$) are also listed. Galactic longitude and latitude are $l$ and $b$, respectively.} \label{tab:star_photo_fesc} \\
\toprule
                Name &   $\mathrm{mag}$&  $ {l}$  &    $ {b}$  & Dist &     S(type) &  S(type) reference &     $r$ &\\
                     &  &  (deg)    & (deg)  & (pc$\pm$5\% )   &       &        & (pc)   &  \\
\midrule
\endfirsthead

\toprule
              Name &   $\mathrm{mag}$  &  $ {l}$  &    $ {b}$  &     Dist &     S(type) &  S(type)reference &     $r$ &      \\
                     &   & (deg)    & (deg)  &  (pc)$\pm$5\%    &       &        & (pc) & \\
\midrule
\endhead
\midrule
\multicolumn{8}{r}{{Continued on next page}} \\
\midrule
\endfoot

\bottomrule
\endlastfoot
LS  IV -15   42 & 10.9 & 15.70 & -0.06 & 2877.7 &           O8.5V+O8.5V & a&         629.8 \\
LS  IV -13    3 & 11.6 & 15.88 &  4.23 & 2125.8 &             ON6V((f)) & a&         195.9 \\
      HD 168444 &  8.8 & 16.20 & -0.01 & 1542.7 &            O9.5/B0Iab & c&         796.5 \\
      V* QR Ser &  9.2 & 16.81 &  0.67 & 2078.1 &     O9.5III+B3-5V/III & h&         307.9 \\
    V* V479 Sct & 12.0 & 16.88 & -1.29 & 2040.8 &            ON6V((f))z & a&         319.0 \\
    BD-14  5040 & 10.8 & 16.90 & -1.12 & 1716.7 &         O5.5V(n)((f)) & a&         614.5 \\
LS  IV -13   14 & 11.0 & 16.91 &  0.85 & 1830.1 &                   O9V & g&         512.6 \\
      HD 168075 &  8.8 & 16.94 &  0.84 & 1682.0 &          O7V(n)((f))z & b&         651.1 \\
    BD-13  4930 &  9.4 & 16.94 &  0.77 & 1786.0 &                 O9.7V & g&         554.6 \\
    BD-13  4928 & 10.1 & 16.97 &  0.82 & 1692.0 &              O9.7IVnn & g&         641.8 \\
      HD 168137 &  9.5 & 16.97 &  0.76 & 1881.1 &                  O8Vz & a&         467.9 \\
    BD-13  4929 & 10.4 & 16.98 &  0.82 & 1703.0 &      O7V+B0.5V+B0.5V: & h&         631.6 \\
    BD-13  4927 & 10.2 & 16.99 &  0.85 & 1758.7 &               O7II(f) & b&         578.7 \\
      HD 168504 & 10.0 & 17.03 &  0.35 & 1717.3 &             O7.5V(n)z & a&         625.1 \\
      HD 167330 &  9.6 & 17.66 &  2.16 & 1728.0 &          O9.5/B0Ia/ab & c&         590.3 \\
      HD 169727 &  9.9 & 17.99 & -0.82 & 2090.3 &                  O7II & d&         293.7 \\
    BD-12  5004 & 11.3 & 18.22 &  1.17 & 1874.7 &                  O9IV & d&         466.5 \\
    BD-12  4979 & 10.4 & 18.25 &  1.69 & 1912.0 &                 O9.5V & e&         423.3 \\
LS  IV -12   12 & 10.4 & 18.32 &  1.87 & 1829.1 &              O6V((f)) & a&         498.1 \\
      HD 168112 &  9.8 & 18.44 &  1.62 & 2006.0 &              O5III(f) & b&         341.8 \\
    BD-12  4984 & 10.0 & 18.53 &  1.77 & 2026.7 &                   O9V & o&         321.4 \\
      HD 168461 & 10.2 & 18.57 &  1.25 & 2125.4 &        O7.5V((f))Nstr & a&         259.3 \\
      HD 171589 &  8.3 & 18.64 & -3.09 & 1848.7 &             O7.5II(f) & b&         466.0 \\
    BD-12  5042 & 11.6 & 18.69 & -0.02 & 2181.0 &                  O9Ib & d&         272.8 \\
      V* RY Sct & 10.4 & 18.71 & -0.13 & 2106.1 &              O9.7Ibep & l&         309.3 \\
      HD 168206 &  9.1 & 18.91 &  1.75 & 2042.4 &         WC8+O8/9III/V & f&         311.3 \\
      HD 166734 &  8.4 & 18.92 &  3.63 & 1759.0 &              O7.5Iabf & b&         550.6 \\
    BD-11  4586 &  9.4 & 19.08 &  2.14 & 2009.6 &               O8Ib(f) & b&         332.6 \\
    BD-11  4620 & 10.2 & 19.15 &  0.67 & 2139.9 &                    O5 & o&         274.8 \\
    BD-12  5039 & 10.8 & 19.17 &  0.32 & 2115.0 &                    O6 & k&         301.4 \\
      HD 171198 &  9.8 & 20.03 & -1.74 & 2173.9 &                   O7: & c&         235.1 \\
    BD-10  4682 &  9.6 & 20.24 &  1.01 & 1979.8 &             O7Vn((f)) & a&         394.3 \\
    BD-11  4674 & 10.2 & 20.98 & -1.22 & 2887.6 &                    O9 & o&         635.3 \\
    HD 175754  & 7.0   & 16.3  & -9.9  & 2181.9 &         O8II(n)((f))p & e &    367.8\\
    HD 175876  & 6.9   & 15.3  &-10.6  & 2463.1 &         O6.5III(n)(f) & e &   240.9\\

\end{longtable}
\begin{minipage}{0.5\textwidth}
\begin{itemize}
\item[a] \cite{2016ApJS..224....4M}
\item[b] \cite{2011ApJS..193...24S}
\item[c] \cite{1988mcts.book.....H}
\item[d]  \cite{1993ApJS...89..293V}
\item[e] \cite{2014ApJS..211...10S} 
\item[f] \cite{2014yCat....1.2023S}
\item[g] \cite{2022AA...657A.131M}
\item[h] \cite{2009MNRAS.400.1479S}
\item[k]   \cite{2003AJ....125.2531R}
\item[o] \cite{1963LS....C04....0N}
\item[l] \cite{1982AJ.....87.1300W} 
\end{itemize}
\end{minipage}

\end{onecolumn}

\begin{onecolumn}
\begin{longtable}{lcccccccc}
\caption{List of OB stars from \citet{2003AJ....125.2531R}. Their spectral type was found using {\tt Astropy} and {\tt SIMBAD}, and the stars are limited to the region between $15.5^\circ \leq l \leq 21^\circ$ and  $-5^\circ \leq b \leq 5^\circ$. The stars' magnitudes (mag) are from \cite{2003AJ....125.2531R} (mag). The distances to Earth (based on Gaia DR3 parallax measurements; \citet{2020yCat.1350....0G}) and the distances $r$ to the bow-shock (GL=17.3$^{\circ}$ and GB=$-$5.5$^{\circ}$) are also listed. Galactic longitude and latitude are $l$ and $b$, respectively.}  \label{tab:star_photo_fesc2} \\
\toprule
                Name &   $\mathrm{mag}$&  $ {l}$  &    $ {b}$  &     Dist &     S(type) &  S(type) reference &     $r$ &\\
                     &  &  (deg)    & (deg)  &  (pc)    &       &        & (pc)   &  \\
\midrule
\endfirsthead

\toprule
              Name &   $\mathrm{mag}$  & $ {l}$  &    $ {b}$  &   Dist &     S(type) &  S(type) reference &    ${r}$ &      \\
                     &   & (deg)    & deg  &  (pc)    &       &        & (pc)  & \\
\midrule
\endhead
\midrule
\multicolumn{8}{r}{{Continued on next page}} \\
\midrule
\endfoot

\bottomrule
\endlastfoot
 LS  IV -12    3 &11.4  &16.23 &  4.88 & 1641.49&           OB- & a &     765.19 \\
     BD-12  4915 & 11.5 &16.81 &  4.35 & 1992.82&            OB & a &     479.4 \\
         LS 4657 & 11.9 &15.57 &  3.49 & 2097.31&           OB+ & a &     400.6 \\
 LS  IV -13    4 & 11.6 &15.60 &  2.93 & 1682.36&           OB- & b &     698.3\\
 LS  IV -14   10 & 11.4 &15.77 &  1.63 & 1751.00&           OB- & b &     616.0\\
     BD-14  4937 & 10.8 &15.60 &  1.49 & 1280.73&           OB- & b &    1056.0\\
     BD-12  4949 & 10.7 &17.17 &  2.32 & 1656.45&            OB & a &     707.6 \\
 TYC 5685-3491-1 & 11.4 &17.89 &  1.93 & 1675.32&            OB & a &     684.1 \\
 LS  IV -13    7 & 11.6 &16.81 &  1.13 & 1680.39&           OB- & b &     669.6\\
     BD-12  4964 & 10.8 &18.11 &  1.80 & 2025.93&            OB & a &     387.6 \\
 LS  IV -14   28 & 11.5 &15.87 &  0.50 & 1520.45&            OB & b &     815.7\\
 LS  IV -12   14 & 11.6 &18.31 &  1.78 & 1920.49&            OB & b &     467.8\\
     BD-12  4975 & 11.1 &18.20 &  1.70 & 1944.39&           OB- & a &     446.8 \\
     BD-12  4977 & 10.7 &18.16 &  1.67 & 2144.08&            OB & b &     311.3\\
     BD-12  4978 & 10.4 &18.24 &  1.70 & 2089.86&            OB & b &     342.7\\
  TYC 5689-418-1 & 10.6 &17.28 &  1.18 & 2055.49&           OB+ & a &     352.1 \\
     BD-12  4982 & 10.1 &18.31 &  1.70 & 2023.88&            OB & a &     386.6 \\
 LS  IV -12   27 & 11.3 &18.39 &  1.73 & 2181.50&            OB & a &     296.7 \\
 LS  IV -14   29 & 11.4 &16.50 &  0.66 & 1631.85&           OB- & b &     709.8\\
 LS  IV -12   32 & 11.0 &17.98 &  1.33 & 1593.62&           OB- & b &     751.9\\
 LS  IV -12   33 & 12.1 &18.48 &  1.59 & 2166.37&           OB+ & a &     298.9 \\
 LS  IV -12   34 & 10.5 &18.58 &  1.60 & 2157.96&           OB- & b &     303.0\\
     BD-12  4994 & 10.6 &18.47 &  1.54 & 2218.27&            OB & a &     277.7 \\
 LS  IV -14   32 & 11.1 &16.29 &  0.34 & 1280.57&           OB- & a &    1045.2 \\
 LS  IV -12   36 & 11.1 &18.01 &  1.25 & 1533.74&           OB- & b &     807.5\\
     BD-12  4996 & 11.5 &18.00 &  1.24 & 2562.13&            OB & a &     365.2 \\
 LS  IV -12   38 & 11.3 &18.09 &  1.25 & 1702.99&           OB- & b &     649.0\\
 LS  IV -12   39 & 11.3 &18.35 &  1.38 & 2198.76&           OB- & a &     278.4 \\
     BD-13  4941 & 10.7 &17.76 &  0.87 & 2099.95&           OB+ & a &     313.5 \\
     BD-12  5008 & 10.8 &18.13 &  1.05 & 1974.33&            OB & a &     409.0 \\
 LS  IV -09    5 & 11.8 &20.87 &  2.44 & 1851.85&           OB- & b &     547.2\\
 LS  IV -14   39 & 11.6 &16.60 & -0.07 & 1940.61&            OB & a &     417.8 \\
 TYC 5689-1271-1 & 11.3 &16.38 & -0.24 & 2351.28&           OB+ & a &     215.5 \\
         LS 4965 & 11.4 &17.18 &  0.12 & 1787.94&           OB- & a &     555.9 \\
  TYC 5689-237-1 & 11.0 &17.20 &  0.07 & 2237.13&           OB+ & a &     223.8 \\
 GSC 06265-01861 & 11.4 &15.53 & -0.82 & 2125.85&            OB & c &     265.7\\
 TYC 6265-2079-1 & 11.0 &15.59 & -0.82 & 1683.21&            OB & c &     648.4\\
 LS  IV -11   16 & 11.9 &19.75 &  1.25 & 1533.74&           OB- & b &     809.8\\
         LS 4992 & 11.2 &17.84 &  0.10 & 1971.99&           OB+ & a &     391.5 \\
 LS  IV -10    7 & 11.7 &20.25 &  1.23 & 1543.92&           OB- & b &     801.5\\
 LS  IV -12   47 & 12.6 &18.88 &  0.44 & 2010.05&           OB- & a &     369.0 \\
     BD-14  5029 & 10.6 &17.23 & -0.44 & 2036.66&            OB & a &     328.5 \\
 LS  IV -13   43 & 11.0 &17.75 & -0.18 & 1560.30&           OB- & a &     767.2 \\
UCAC4 380-107183 & 11.4 &17.25 & -0.46 & 2614.37&           OB- & a &     370.3 \\
 TYC 5702-1189-1 & 11.4 &16.54 & -0.90 & 2040.81&           OB+ & a &     318.7 \\
 LS  IV -14   49 & 11.6 &17.46 & -0.52 & 2297.79&            OB & a &     192.5 \\
 LS  IV -14   50 & 11.7 &16.91 & -0.84 & 1634.78&           OB- & b &     690.3\\
 LS  IV -14   51 & 11.8 &17.49 & -0.62 & 1749.78&           OB- & a &     581.8 \\
         LS 5016 & 11.7 &17.62 & -0.57 & 2554.27&          OB+: & a &     317.7 \\
         LS 5022 & 11.1 &16.10 & -1.49 & 2158.42&           OB+ & a &     222.0 \\
 TYC 5702-1096-1 & 11.4 &18.05 & -0.52 & 1963.09&            OB & a &     387.7 \\
 LS  IV -12   55 & 11.4 &18.68 & -0.19 & 2482.00&           OB: & a &     274.0 \\
UCAC4 384-101260 & 11.5 &18.20 & -0.45 & 2868.61&            OB & a &     600.8 \\
 TYC 5698-3744-1 & 11.5 &18.63 & -0.27 & 2139.03&           OB+ & a &     258.1 \\
     BD-14  5043 & 10.3 &17.06 & -1.15 & 1179.80&            OB & a &    1133.4 \\
 LS  IV -13   45 & 11.4 &18.18 & -0.57 & 2341.92&           OB- & a &     194.6 \\
 LS  IV -15   46 & 10.8 &16.51 & -1.48 & 2675.94&            OB & a &     411.0 \\
 LS  IV -11   19 & 12.1 &20.12 &  0.38 & 2034.58&          OB-e & b &     355.1\\
 LS  IV -15   49 & 11.3 &16.65 & -1.45 & 1906.94&           OB- & b &     426.6\\
 LS  IV -13   46 & 11.6 &17.70 & -0.95 & 2014.91&           OB- & a &     335.5 \\
     BD-14  5048 & 11.6 &16.87 & -1.42 & 2308.40&           OB- & a &     164.9 \\
 LS  IV -14   63 & 11.1 &17.47 & -1.20 & 2302.55&           OB- & a &     168.9 \\
 TYC 5698-4048-1 & 11.4 &18.98 & -0.44 & 2121.79&            OB & a &     266.0 \\
     BD-14  5054 & 10.1 &17.10 & -1.44 & 1591.84&           OB- & b &     726.5\\
     BD-14  5058 &  9.5 &17.77 & -1.15 & 1922.33&           OB- & a &     414.1 \\
 LS  IV -14   71 & 11.4 &17.02 & -1.62 & 2039.15&           OB- & b &     304.4\\
       HD 170061 &  9.9 &17.20 & -1.60 & 1874.76&           OBe & a &     453.1 \\
     BD-13  4990 & 11.0 &18.66 & -0.91 & 2053.81&           OB- & a &     304.7 \\
     BD-16  4894 & 10.4 &15.65 & -2.72 & 1979.41&            OB & a &     350.2 \\
 LS  IV -15   53 & 11.1 &17.05 & -2.01 & 1951.98&           OB- & b &     376.3\\
 LS  IV -12   68 & 11.4 &19.09 & -1.23 & 2331.54&           OB- & a &     173.9 \\
 LS  IV -11   22 & 11.9 &20.39 & -0.61 & 1044.38&           OB- & b &    1272.5\\
     BD-11  4652 & 11.5 &20.27 & -0.77 & 1816.86&           OB- & b &     522.8\\
 LS  IV -12   73 & 12.0 &19.29 & -1.31 & 1719.69&           OB- & b &     604.7\\
 LS  IV -13   59 & 11.7 &18.89 & -1.52 & 2402.11&           OB- & b &     189.6\\
 LS  IV -13   60 & 12.0 &19.02 & -1.48 & 2288.85&           OB- & b &     162.3\\
 LS  IV -13   61 & 11.5 &18.46 & -1.85 & 1817.85&           OB- & b &     503.7\\
 LS  IV -13   67 & 11.2 &18.93 & -1.84 & 1785.39&           OB- & b &     535.7\\
     BD-12  5095 & 10.4 &19.42 & -1.73 & 2146.84&           OB- & a &     219.6 \\
 LS  IV -13   68 & 11.1 &18.85 & -2.05 & 2194.42&           OB- & b &     176.3\\
 LS  IV -12   80 & 11.6 &19.45 & -1.80 & 2602.81&           OB- & b &     340.2\\
     EM* AS  309 & 11.2 &18.54 & -2.42 & 1885.37&          OB-e & b &     433.5\\
 LS  IV -12   83 & 11.7 &19.73 & -1.81 & 2338.08&           OB- & a &     163.1 \\
 LS  IV -13   72 & 10.8 &18.76 & -2.33 & 2160.76&           OB- & a &     191.2 \\
     BD-12  5103 & 11.0 &20.15 & -1.66 & 1459.64&           OB- & b &     857.2\\
 LS  IV -13   74 & 12.0 &18.99 & -2.30 & 1950.07&           OB- & b &     374.6\\
 LS  IV -14   88 & 11.7 &18.25 & -2.70 & 1886.43&           OB- & b &     429.4\\
 LS  IV -14   89 & 11.9 &18.28 & -2.73 & 2173.44&           OB- & b &     170.7\\
     BD-13  5035 & 11.2 &18.88 & -2.55 & 2097.75&           OB- & b &     237.4\\
 LS  IV -12   92 & 11.8 &19.91 & -2.07 & 1301.23&           OB- & b &    1010.4\\
 LS  IV -14   91 & 12.0 &18.45 & -2.83 & 2151.00&           OB- & b &     186.1\\
 LS  IV -11   31 & 11.8 &20.49 & -1.88 & 1479.72&           OB- & b &     837.3\\
 LS  IV -14   94 & 11.9 &18.27 & -3.13 & 2642.70&           OB- & b &     357.2\\
 LS  IV -13   80 & 11.5 &19.63 & -2.80 & 1924.18&           OB- & b &     396.9\\
 LS  IV -14   96 & 11.7 &18.91 & -3.32 & 1751.92&           OB- & b &     556.9\\
 LS  IV -12   96 & 11.6 &19.85 & -2.87 & 1076.77&           OB- & b &    1229.8\\
 LS  IV -11   33 & 11.4 &20.95 & -2.46 & 1751.92&           OB- & b &     571.2\\
 TYC 6271-1204-1 & 11.8 &16.41 & -4.80 & 2962.08&           OB- & a &     665.7 \\
     BD-12  5133 & 11.3 &20.59 & -3.13 & 2794.85&           OB- & a &     513.0 \\
     BD-13  5065 & 10.1 &19.74 & -3.67 & 1872.30&         OB-n: & a &     439.5 \\
         LS 5116 & 12.2 &19.38 & -4.13 & 1787.63&            OB & a &     518.8 \\
 LS  IV -13   86 & 11.7 &19.94 & -3.94 & 1755.92&           OB- & b &     553.0\\
 LS  IV -14  102 & 11.2 &19.06 & -4.69 & 1678.41&           OB- & b &     624.3\\
\end{longtable}
\begin{minipage}{0.5\textwidth}
\begin{itemize}
\item[a] \cite{1971PWSO...1a...1S}
\item[b] \cite{1963LS....C04....0N}
\item[c] \cite{2003AJ....125.2531R}
\end{itemize}
\end{minipage}
\end{onecolumn}

\end{document}